\documentstyle[preprint,pre,aps,epsf,eqsecnum]{revtex}
\tighten    

\begin{document}
\draft   
\preprint{Submitted to Physical Review E.}

\title{Hypersensitivity to perturbation in the quantum kicked top}

\author{R\"udiger Schack$^{^{\hbox{\tiny(a)}}}$,
        Giacomo M. D'Ariano$^{^{\hbox{\tiny(b)}}}$,
  and   Carlton M. Caves$^{^{\hbox{\tiny(a)}}}$}

\address{$^{\hbox{\tiny(a)}}$Center for Advanced Studies,
Department of Physics and Astronomy,\\
University of New Mexico,
Albuquerque, NM~87131-1156, USA\\
$^{\hbox{\tiny(b)}}$Dipartimento di Fisica ``Alessandro Volta,''
Universit\`a degli Studi di Pavia,\\
Via A.~Bassi 6, 27100 Pavia, Italy}

\date{March 25, 1994}
\maketitle

\begin{abstract}
For the quantum kicked top we study numerically the distribution of
Hilbert-space vectors evolving in the presence of a small random
perturbation. For an initial coherent state centered in a
chaotic region of the classical dynamics, the evolved perturbed
vectors are distributed essentially like random vectors in Hilbert
space. In contrast, for an initial coherent state centered near an
elliptic (regular) fixed point of the classical dynamics, the evolved
perturbed vectors remain close together, explore only few dimensions
of Hilbert space, and do not explore them randomly. These results
support and extend results of earlier studies, thereby providing
additional support for a new characterization of quantum chaos that
uses concepts from information theory.
\end{abstract}

\narrowtext

\section{INTRODUCTION}    

In a series of previous
papers\cite{Schack1993e,Caves1993b,Caves1993a,Schack1992a}, two of the
authors introduced a characterization of Hamiltonian chaos which
is directly applicable to quantum as well as to classical systems.
This characterization, formulated in the framework of statistical
mechanics, is based on the following question: How much information
is needed to predict an evolved system state in the presence of random
perturbations?  General arguments\cite{Caves1993b} and an investigation
of the symbolic dynamics of the baker's map\cite{Schack1992a} provide
strong evidence that chaotic classical Hamiltonian systems show what
we call {\it hypersensitivity to perturbation}---i.e., a rapid increase
with the number of time steps of the information needed to describe the
perturbed time evolution of a system state, the information attaining
values exponentially larger than the increase of ordinary entropy that
results from averaging over the perturbation.

Hypersensitivity to perturbation explains quantitatively how
information about the state of a system is lost through interaction
with an incompletely known environment and, therefore, is important for
understanding why entropy necessarily increases in systems that are
not perfectly isolated. The connection of this work on chaos with
statistical physics is developed in Section~\ref{STAT}.  In addition
to its straightforward motivation in statistical mechanics,
the concept of hypersensitivity to perturbation may provide a more
physical way to characterize quantum chaos\cite{Caves1993b}. Numerical
simulations\cite{Schack1993e} show that the quantum baker's
map\cite{Balazs1989} displays hypersensitivity to perturbation.

In this paper, we analyze how hypersensitivity to perturbation arises
in the quantum kicked top\cite{Frahm1985,Haake1987}, a system whose
classical dynamics has both chaotic and regular regions. To shed light
on the reason for the rapid increase of information associated with
the property of hypersensitivity to perturbation, we perform a
detailed numerical analysis of how the vectors arising from different
perturbation histories (realizations of the random perturbation) are
distributed in Hilbert space. For an initial coherent state centered
in a chaotic region of the classical dynamics, the evolved perturbed
vectors are distributed essentially like random vectors in Hilbert
space. In contrast, for an initial coherent state centered near an
elliptic (regular) fixed point of the classical dynamics, the evolved
perturbed vectors remain close together, explore only few dimensions
of Hilbert space, and do not explore those dimensions randomly.

Quantum systems show no ``sensitivity to initial conditions,'' due to
unitarity, but they show what one might call sensitivity to parameters
in the Hamiltonian, as has been demonstrated for the kicked top by
Peres\cite{Peres1991b}. Peres compares the time evolution of the same
initial Hilbert-space vector for two slightly different values of the
twist parameter in the kicked-top Hamiltonian (see Section~\ref{TOP}).
He finds that, after a fixed number of time steps, the two evolved
vectors are far apart if the initial vector is a coherent state centered
in a chaotic region of the classical dynamics, but the two evolved
vectors stay close together if the initial coherent state is centered
near an elliptic fixed point of the classical dynamics.

Our approach to quantum chaos could be viewed as a
generalization of Peres's work: while Peres studies time evolution due
to an incompletely known Hamiltonian, we analyze the distribution of
vectors arising from time evolution under a stochastic Hamiltonian.
There is, however, a fundamental difference in philosophy between the
two approaches, which can be understood fully only in the context of
statistical mechanics. This difference becomes apparent in
Section~\ref{STAT}.

The quantum kicked top, its classical limit, and the concept of a
coherent state are reviewed in Section~\ref{TOP}.  Section~\ref{PERT}
defines the perturbations used in the numerical simulations. In
Section~\ref{DIST} we explain how the distribution of vectors in
Hilbert space is connected to questions of information and entropy and
how the numerical data are compiled into figures.  Section~\ref{NUM}
contains the numerical results of this paper.  Finally, in
Section~\ref{STAT}, we discuss the implications of our results for the
foundations of statistical physics.

\section{THE KICKED TOP} \label{TOP}

The quantum model of the kicked top\cite{Frahm1985,Haake1987}
describes a spin-$J$ particle---i.e., an angular momentum vector
$\hbar
\hat{\bf J}=\hbar ({\hat J}_x,{\hat J}_y,{\hat J}_z)$,
where $[{\hat J}_i ,{\hat J}_j ]=i\epsilon_{ijk}{\hat J}_k$---whose
dynamics in $(2J+1)$-dimensional Hilbert space is governed by the
Hamiltonian

\begin{eqnarray}
\hat H (t)=(\hbar p/T){\hat J}_z+(\hbar k/{2J}){\hat J}_x^2
\sum_{n=-\infty}^{+\infty}{\delta(t-nT)}\;.
\label{H}
\end{eqnarray}
The free precession of the spin around the $z$ axis (first term in
the Hamiltonian) is interrupted periodically by sudden kicks or {\it
twists\/} at times $nT$ with twist parameter $k$ (second term in the
Hamiltonian). The angle of free precession between kicks is given by
$p$. In this paper we always use $p=\pi/2$.

We look at the time evolution of an initial Hilbert-space vector
$|\psi_0\rangle$ at discrete times $nT$. After $n$ time steps, the
evolved vector is given by
\begin{eqnarray}
|\psi_n\rangle =\hat U_k^n |\psi_0\rangle\;,
\end{eqnarray}
where $\hat U_k$ is the unitary Floquet operator:
\begin{eqnarray}
\hat U_k=\exp{[-i(k/2J){\hat J}_x^2]}\exp{(-i\pi{\hat J}_z/2)}\;.
\label{U}
\end{eqnarray}

The classical Poincar\'e map corresponding to the quantum map is
obtained by introducing the unit vector $\vec\omega =(X,Y,Z) \equiv
\hat{\bf J}/J$ and performing the limit $J\to\infty$. One
obtains\cite{Haake1987}
\begin{eqnarray}
&&X'=-Y \nonumber\;,\\
&&Y'=X\cos k Y +Z\sin k Y\;,\label{Poincare} \\
&&Z'=Z\cos k Y -X\sin k Y\;.\nonumber
\end{eqnarray}
The map (\ref{Poincare}) is an area-preserving map of the unit sphere,
i.e., an area-preserving map on the configuration space of a classical
spin with fixed magnitude. Depending on the value of the twist
parameter $k$, this map has regions of chaotic behavior interspersed
with regular regions associated with elliptic cyclic points. In this
paper, we are interested in two cyclic points of the map for $k=3$.
One is an elliptic fixed point of period 1 located at\cite{D'Ariano1992}
\begin{equation}
Z=\cos\theta=0.455719\;,\;\;\varphi=3\pi/4\;,
\label{elliptic}
\end{equation}
where we have used spherical coordinates
\begin{eqnarray}
\theta &= &\arccos Z\;, \nonumber\\
\varphi   &= &\arctan Y/X\;.
\label{spherical}
\end{eqnarray}
The elliptic fixed point (\ref{elliptic}) is surrounded by an
oval-shaped regular region, extending about $0.3$ radians in the
$\varphi$-direction and about $0.5$ radians in the
$\theta$-direction. The other cyclic point of interest to us is a
hyperbolic fixed point of period 4, which has a positive Lyapunov
exponent. It is located in the middle of a chaotic region
at\cite{D'Ariano1992}
\begin{equation}
Z=\cos\theta=0.615950\;,\;\;\varphi=\pi/4\;.
\label{hyperbolic}
\end{equation}
We choose the two fixed points (\ref{elliptic}) and (\ref{hyperbolic})
instead of the extreme elliptic ($\varphi=n\pi/4$, $Z=0$) and
hyperbolic ($Z=\pm1$) points described in
Ref.~\cite{D'Ariano1992}, because the latter suffer accidental
invariance with respect to one or both of the perturbation operators
considered in Section~\ref{PERT}.

The model has a conserved parity, which for half-integer $J$ takes the
form
\begin{equation}
\hat S =-i\exp(-i\pi\hat J_z)\label{parity}
\label{parityS}
\end{equation}
and which permits factorization of the matrix representation of the
operator $\hat U$ into two blocks. Starting from a state with definite
parity, the whole dynamical evolution occurs in the invariant
Hilbert subspace with the given parity.  For half-integer $J$, the
dimension of the even-parity subspace (eigenspace of $\hat S$ with
eigenvalue 1) is $J+\case1/2$. In this paper, we work with $J=511.5$
in the 512-dimensional even-parity subspace. We consider only the
projection of the initial vector in the even subspace.  Numerical
evidence and symmetry considerations\cite{Peres1991b} suggest that no
additional insight is gained by including the odd-parity subspace.  In
any case, the restricted model can be regarded as a quantum map in its
own right, which can be investigated independently of the behavior of
the complete kicked-top model.

We want to choose initial vectors for the quantum evolution that
correspond as closely as possible to the classical directions
(\ref{elliptic}) and (\ref{hyperbolic}). For this purpose, coherent
states\cite{Radcliffe1971,Atkins1971,Perelomov1986} are appropriate.
The coherent state
$|\theta,\varphi\rangle$ is defined by the relation
\begin{equation}
{\bf n}\cdot\hat{\bf J}|\theta,\varphi\rangle=J|\theta,\varphi\rangle\;,
\end{equation}
where ${\bf n}$ is the unit vector pointing in the direction given by
$\theta$ and $\varphi$. All coherent states can be generated by an
appropriate rotation of the state
$|J,M=J\rangle=|\theta=\pi/2,\varphi=0\rangle$, where $|J,M\rangle$
($M=-J,\ldots,J$) is the common eigenstate of $\hat J^2$ and $\hat J_z$
with eigenvalues $J(J+1)$ and $M$, respectively. In
calculations, it is convenient to use the explicit representation
\begin{equation}
|\theta,\varphi\rangle =\sum_{n=0}^{2J}{\sqrt{P_J (\theta,n)}
e^{in\varphi}|J,n-J\rangle }\;,
\label{i-phi}
\end{equation}
where
\begin{equation}
P_J (\theta,n)={2J\choose n}\left( {1+\arccos\theta\over2}\right)
^{2J-n}\left({1-\arccos\theta\over2}\right)^n\;.
\label{binomial}
\end{equation}

In the following we need a metric on Hilbert space. The distance
between two normalized vectors $|\psi_1\rangle$ and $|\psi_2\rangle$
is defined as the Hilbert-space angle
\begin{equation}
s(|\psi_1\rangle,|\psi_2\rangle)
=\cos^{-1}(|\langle\psi_1|\psi_2\rangle|)\equiv\phi
\label{Hdistance}
\end{equation}
 between the two
vectors\cite{Wootters1981}.  Consider two coherent states
$|\theta,\varphi\rangle$ and $|\theta',\varphi'\rangle$. In terms of
the angle $\alpha$ between the directions $(\theta,\varphi)$ and
$(\theta',\varphi')$, the distance between the two coherent states is
given by\cite{Perelomov1972}
\begin{equation}
\cos[s(|\theta,\varphi\rangle,|\theta',\varphi'\rangle)]=
|\langle\theta,\varphi\mid\theta',\varphi'\rangle|=
[\cos(\alpha/2)]^{2J}\simeq\exp(-J\alpha^2/4)\;,
\end{equation}
where the approximation is valid for large $J$. Two coherent states
can therefore be regarded as roughly orthogonal if
$\alpha\gtrsim2J^{-1/2}$\cite{Peres1991b}.  The {\it size\/} of the
coherent state $|\theta,\varphi\rangle$ is conveniently defined in
terms of the $Q$ function
\begin{equation}
Q_{\theta,\varphi}(\theta',\varphi')\equiv
|\langle\theta',\varphi'\mid\theta,\varphi\rangle|^2=
[\cos(\alpha/2)]^{4J}\equiv Q(\alpha)\;.
\end{equation}
Since $Q(2J^{-1/2})\simeq e^{-2}Q(0)$, the $Q$ function of the
coherent state $|\theta,\varphi\rangle$ is very small outside a region
of radius $2J^{-1/2}$ centered at the direction $(\theta,\varphi)$.
For the value $J=511.5$ used in this paper, one finds a radius of
$2J^{-1/2}\simeq0.09$ radians, less than the size of the regular
region around the elliptic fixed point (\ref{elliptic}).

\section{PERTURBED EVOLUTION} \label{PERT}

Our goal is to quantify how much information is required to
track the state of a system when, instead of being wholly isolated, it
is perturbed by interaction with its environment.  In classical physics,
interactions with an incompletely known environment can be described
by a stochastic Hamiltonian, each realization of which corresponds
to a particular initial condition for the environment. The situation is
more complicated in quantum mechanics.  Due to the possible entanglement
of system and environment, there is no way to associate a unique
perturbation history with a given initial condition of the environment.
An upcoming publication discusses how to study hypersensitivity to
perturbation in a realistic model of a quantum system interacting with
an environment. For the present paper we restrict ourselves to the
special case of quantum time evolution under a stochastic Hamiltonian.

The problem is simplified further by considering only
two possible unitary time evolutions at each step. These two time
evolution operators we denote by $\hat U^+$ and $\hat U^-$. A
perturbed time step consists in applying either $\hat U^+$ or $\hat
U^-$ with equal probability. After $n$ time steps, the number of
different perturbation sequences is $2^n$, each sequence having
probability $2^{-n}$.

We use two different perturbations: (i) the {\it twist perturbation},
defined by choosing the twist parameter $k$ at random from step to
step, the two possible Floquet operators being
given by\cite{Peres1991b}
\begin{equation}
\hat U^+=\hat U_k\;,\;\;\;
\hat U^-=\hat U_{k+\epsilon}\;;
\label{twist}
\end{equation}
(ii) the {\it turn perturbation}, defined by rotating the spin by a
small angle $\epsilon$ around the $z$ axis after each unperturbed step
$\hat U_k$, the two possible Floquet operators being given by
\begin{equation}
\hat U^+=\exp(-i\epsilon\hat J_z)\,\hat U_k\;,\;\;\;
\hat U^-=\exp(+i\epsilon\hat J_z)\,\hat U_k\;.
\label{turn}
\end{equation}
Notice that the time-evolution operators~(\ref{twist}) and (\ref{turn})
commute with parity (\ref{parity}) and hence do not couple odd- and
even-parity subspaces.

For perturbation  strengths we use $\epsilon=0.03$
and $\epsilon=0.003$ for the twist perturbation and $\epsilon=0.003$
for the turn perturbation. For a twist parameter $k+\epsilon$ the
zenithal location of the fixed points (\ref{elliptic}) and
(\ref{hyperbolic}) changes slightly to $Z=0.443579$ and $Z=0.619848$
for $\epsilon=0.03$, and to $Z=0.454497$ and $Z=0.616341$ for
$\epsilon=0.003$. This corresponds to changes in the zenithal angle
$\theta$ of $\Delta\theta\simeq0.014$ radians and
$\Delta\theta\simeq0.005$ radians for $\epsilon=0.03$, and to
$\Delta\theta\simeq0.0014$ radians and $\Delta\theta\simeq0.0005$
radians for $\epsilon=0.003$. All these angles, as well as the angle
$\epsilon=0.003$ we use for the turn perturbation, are very small
compared to the size of the elliptic region around the fixed point
(\ref{elliptic}) and are also much smaller than the size of a coherent
state.

\section{DISTRIBUTION OF VECTORS AND INFORMATION} \label{DIST}

The $2^n$ different perturbation sequences obtained by applying every
possible sequence of $\hat U^-$ and $\hat U^+$ for $n$ time steps lead
to a list of $2^n$ vectors, each having probability $2^{-n}$.  In this
section, we explain how the distribution of these $2^n$ vectors in
Hilbert space is related to information and entropy.

Let us start with a slightly more general situation. Imagine we are
given a list of $N$ vectors in $D$-dimensional Hilbert space,
$|\psi_1\rangle,\ldots,|\psi_N\rangle$, with probabilities
$p_1,\ldots,p_N$. Together with our knowledge of the system
Hamiltonian and boundary conditions, the list of vectors with their
probabilities constitutes our {\it background information}. We ask for
the average information needed to specify a single one of these
vectors, given the background information. The information to specify
a particular vector can be quantified either via conditional
algorithmic information\cite{Chaitin1987a} or by the length of a
codeword in some coding scheme\cite{Gallager1968}. In both cases, it
is a consequence of the variable-length coding
theorem\cite{Gallager1968} that the information averaged over all
vectors, or {\it average information\/}, is bounded below by
\begin{equation}
\Delta I=-\sum_{i=1}^N p_i\log_2 p_i\;.
\end{equation}
(Throughout this paper, information and entropy are measured in
bits.)  There exist coding schemes---an
example is Huffman coding\cite{Huffman1952}---where $\Delta I+1$ is an
upper bound for the average information or codeword length. It can be
shown\cite{Schack1994c} that there exists a universal computer for
which $\Delta I+1$ is an upper bound for the average algorithmic
information as well. Therefore, we call $\Delta I$ the average
information, in the sense that the actual average information is
within one bit of $\Delta I$ if an efficient coding scheme is used. In
the case of $2^n$ equiprobable vectors the average information is
$\Delta I=n$.

Suppose some of the $N$ vectors $|\psi_i\rangle$ are very close
together in Hilbert space, so that they form a small group. If one
is interested in lowering the amount of information $\Delta I$,
one may choose to provide just enough information to specify that
the actual vector is located in that group, the price being that the
entropy of the group is generally bigger than zero.  To be more specific,
we introduce a coarse graining on Hilbert space defined by a resolution
angle $\phi$. Vectors less than an angle $\phi$ apart are grouped together.
More precisely, groups are formed in the following way. Starting with
the first vector in the list, $|\psi_1\rangle$, the first group is
formed of $|\psi_1\rangle$ and of all vectors in the list that are
within an angle $\phi$ of $|\psi_1\rangle$. The same procedure is then
repeated with the remaining vectors to form the second group, then the
third group, continuing until no ungrouped vectors are left.
This grouping of vectors corresponds to a partial averaging over the
perturbations. To describe a vector at resolution level $\phi$ amounts
to averaging over those details of the perturbation that do not change
the final vector by more than an angle $\phi$.

For a given resolution $\phi$, there are $N(\phi)$ groups. We
denote by $N_j$ the number of vectors in the $j$th group
($\sum_{j=1}^{N(\phi)}N_j=N$). The $N_j$ vectors in the $j$th group
and their probabilities are denoted by
$|\psi^j_1\rangle,\ldots,|\psi^j_{N_j}\rangle$ and
$p^j_1,\ldots,p^j_{N_j}$, respectively. Knowing that the system
state is in the $j$th group, but not knowing which state in the $j$th
group is the actual state, corresponds to describing the system by
the density operator
\begin{equation}
\hat\rho_j=
q_j^{-1}\sum_{i=1}^{N_j}p_i^j|\psi^j_i\rangle\langle\psi^j_i|\;,
\end{equation}
where
\begin{equation}
q_j=\sum_{i=1}^{N_j}p^j_i
\end{equation}
is the probability of the $j$th group given only the background
information. For $\phi=\pi/2$, there is only one group, whose density
operator, denoted by $\hat\rho(\pi/2)$, corresponds to a complete average
over the perturbations.

The average information needed to specify which group a vector is
in---i.e., the average information needed to specify the system state
at resolution level $\phi$---is given by
\begin{equation}
\Delta I(\phi)=-\sum_{j=1}^{N(\phi)}q_j\log_2 q_j\;.
\end{equation}
The entropy of the $j$th group is given by the von Neumann entropy
\begin{equation}
\Delta H^j=-{\rm Tr}[\hat\rho_j\log_2\hat\rho_j]\;,
\end{equation}
and the average entropy, called {\it trade-off entropy\/} in the
following, is
\begin{equation}
\Delta H(\phi)=\sum_{j=1}^{N(\phi)}q_j\Delta H^j\;.
\end{equation}
If one chooses to describe the set of vectors not exactly, but only at
resolution level $\phi$, the average information needed to specify the
system state decreases. There is, however, a price: with increasing
resolution angle, the uncertainty about the system state increases on
the average, to a degree quantified by the trade-off entropy.

As a further characterization of the way the vectors are distributed
in Hilbert space, we want to define a quantity that indicates how many
dimensions of Hilbert space are explored by the vectors in a group.
One such quantity would be the dimension of the subspace spanned by
the vectors, which is equal to the number of nonzero eigenvalues of
$\hat\rho_j$. In practice, however, this is not a very useful measure
because it cannot discriminate between the case in which all
dimensions are occupied with equal weight (all eigenvalues of $\hat\rho_j$
roughly equal) and the case in which most vectors are concentrated in
a low-dimensional subspace (all eigenvalues of $\hat\rho_j$ nonzero, but
of strongly varying magnitude).

A possible measure of the number of explored dimensions, which takes into
account the small weight of dimensions corresponding to relatively
small eigenvalues, is the exponential of the entropy, $2^{\Delta H^j}$.
This quantity is bounded above by $D_j$ if the vectors are confined to
a $D_j$-dimensional subspace and gets smaller if the dimensions are
occupied with different weights. For example, if two eigenvalues of
$\hat\rho_j$ are close to $\case1/2$ and all the others are close to zero,
then $2^{\Delta H^j}\simeq2$, indicating that the vectors are essentially
confined to a two-dimensional subspace.  Unfortunately, for small
resolution angles $\phi$, $\Delta H^j$ is necessarily small just
because all the vectors in the group lie along roughly the same
direction in Hilbert space; this is true even if the orthogonal
components of the vectors are evenly distributed over {\it all\/}
the orthogonal directions in Hilbert space. For example, the density
operator describing a uniform distribution of vectors within a sphere
of radius $\phi\ll\pi/2$ has one dominating eigenvalue close to 1 and
$D-1$ eigenvalues that are all equal and close to zero (see
Appendix~\ref{AEnt}). Clearly, in this case $2^{\Delta H^j}$ is not
an adequate measure of the number of dimensions explored. On the other
hand, if one could disregard the largest eigenvalue in this example,
then the exponential of the entropy would still be a useful measure of
the number of explored dimensions.

We therefore introduce the {\it spread\/} $\Delta H_2^j$ as the
entropy calculated with the largest eigenvalue of $\hat\rho_j$
omitted. The spread is defined as
\begin{equation}
\Delta H_2^j\equiv-\sum_{k=2}^{D_j}{\lambda_k^j\over1-\lambda_1^j}
\log_2\!\left({\lambda_k^j\over1-\lambda_1^j}\right)\;,
\label{defspread}
\end{equation}
where $\lambda_1^j\geq\lambda_2^j\geq\ldots\geq\lambda_{D_j}^j$ are
the nonzero eigenvalues of the density operator $\hat\rho_j$. The average
spread is
\begin{equation}
\Delta H_2(\phi)=\sum_{j=1}^{N(\phi)} q_j\Delta H_2^j\;.
\end{equation}
For a given resolution angle $\phi$, the entropy $\Delta H^j$ is bounded
above by the entropy $H_{D,{\rm max}}(\phi)$ of Eq.~(\ref{HDmax}), which
for small $\phi$ has the value
$H_{D,{\rm max}}(\phi)\simeq\phi^2\log_2[e(D-1)/\phi^2]$.  The spread
$\Delta H_2^j$, on the other hand, can attain its maximum value
$\Delta H_2^j=\log_2(D-1)$ for arbitrary resolution angles $\phi$.
Indeed, the entropy and the spread are related by
\begin{equation}
\Delta H^j=-\lambda_1^j\log_2\lambda_1^j
-(1-\lambda_1^j)\log_2(1-\lambda_1^j)+(1-\lambda_1^j)\Delta H_2^j\;,
\label{entropyspread}
\end{equation}
which indicates how, when there is one dominating eigenvalue, a large
spread does not lead to a large entropy.

By giving different weight to dimensions corresponding to different
eigenvalues of $\hat\rho_j$, the quantity $\lceil2^{\Delta H_2^j}\rceil$
turns out to be a good indicator of the number of Hilbert-space
dimensions explored by the vectors in a group, independent of the size
of the region occupied by the group. ($\lceil x\rceil$ denotes the
smallest integer greater than or equal to $x$.) In our analysis of the
numerical results, we identify the number of dimensions explored
by the total set of $N$ vectors with the integer
$n_d\equiv\lceil2^{\Delta H_2(\pi/2)}\rceil$.

By determining the information $\Delta I(\phi)$, the trade-off entropy
$\Delta H(\phi)$, and the average spread $\Delta H_2(\phi)$ as
functions of the resolution angle $\phi$, a rather detailed picture
emerges of how the vectors are distributed in Hilbert space.  The
information summarizes the distribution of group sizes at the given
resolution.  The trade-off entropy and the average spread indicate how
the vectors are distributed inside the groups.

It is easy to see that information and trade-off entropy obey the
inequalities
\begin{equation}
\Delta I(0)\geq\Delta I(\phi)\geq\Delta I(\pi/2)=0
\label{IneqI}
\end{equation}
and
\begin{equation}
0=\Delta H(0)\leq\Delta H(\phi)\leq\Delta H(\pi/2)\;.
\label{IneqH}
\end{equation}
The first inequality in (\ref{IneqI}) follows from the fact that any
group at resolution $\phi$ is the union of groups at resolution $\phi=0$;
in words, the average information needed to specify a group at resolution
$\phi=0$ is equal to the average information needed to specify a group at
resolution $\phi$ {\it plus\/} the average information needed specify
$\phi=0$ groups within the groups at resolution $\phi$.  The last
inequality in Eq.~(\ref{IneqH}) is a consequence of the concavity of the
von Neumann entropy \cite{Balian1991,Caves1994a}. A general theorem about
average density operators\cite{Balian1991,Caves1994a} shows that, for all
$\phi$,
\begin{equation}
\Delta I(\phi)+\Delta H(\phi)\geq\Delta H(\pi/2)\;. \label{Balian}
\end{equation}
In general, $\Delta I(\phi)$ is a decreasing function of $\phi$,
whereas $\Delta H(\phi)$ is increasing.  This monotonicity can
sometimes be violated, however, because of discontinuous changes in
the grouping of vectors.

As a still further characterization of our list of vectors, we calculate
the distribution $g(\phi)$ of Hilbert-space angles
$\phi=s(|\psi\rangle,|\psi'\rangle)
=\cos^{-1}(|\langle\psi|\psi'\rangle|)$ between all pairs of vectors
$|\psi\rangle$ and $|\psi'\rangle$. For vectors distributed randomly
in $D$-dimensional Hilbert space, the distribution function $g(\phi)$
is computed in Appendix~\ref{AHil}:
\begin{equation}
g(\phi)={d{\cal V}_D(\phi)/d\phi\over{\cal V}_D}=
2(D-1)(\sin\phi)^{2D-3}\cos\phi
\end{equation}
[Eq.~(\ref{gphi})].  Here ${\cal V}_D(\phi)=(\sin\phi)^{2(D-1)}{\cal V}_D$
[Eq.~(\ref{VDPhi})] is the volume contained within a sphere of radius $\phi$
in $D$-dimensional Hilbert space, and ${\cal V}_D=\pi^{D-1}/(D-1)!$
[Eq.~(\ref{VD})] is the total volume of Hilbert space.  The maximum of
$g(\phi)$ is located at $\phi=\arctan(\sqrt{2D-3})$; for large-dimensional
Hilbert spaces, $g(\phi)$ is very strongly peaked near the maximum, which
is located at $\phi\simeq\pi/2-1/\sqrt{2D-3}$, very near $\pi/2$ (see
Fig.~\ref{random}).

\section{NUMERICAL RESULTS} \label{NUM}

In this section, we describe the numerical results, which are shown
in the figures.  In all numerical examples, we use spin $J=511.5$ and
unperturbed twist parameter $k=3$. The calculations are done in the
512-dimensional even-parity subspace (eigenspace of the
parity~(\ref{parityS}) with eigenvalue 1), i.e., effectively
in a 512-dimensional Hilbert space.  Throughout this section,
we use only two different initial
states. The first one, the coherent state $|\theta,\varphi\rangle$
with $\theta$ and $\varphi$ given by Eq.~(\ref{elliptic}), is centered
in a regular region of the classical dynamics; we refer to it as
the {\it regular initial state}. The second one, referred to as
the {\it chaotic initial state}, is the coherent state
$|\theta,\varphi\rangle$ with $\theta$ and $\varphi$ given by
Eq.~(\ref{hyperbolic}); the chaotic initial state is centered in a
chaotic region of the classical dynamics.  Subsections~\ref{TWIST1}
and \ref{TWIST2} describe results for the twist perturbation with two
different perturbations strengths $\epsilon$.  Subsection~\ref{TURN}
contains results for the turn perturbation.

\subsection{Twist perturbation: $\epsilon=0.03$} \label{TWIST1}

Applying all possible $n$-step perturbation sequences, i.e., all
possible sequences of $n$ Floquet operators $\hat U^-$ and $\hat U^+$,
to an initial state $|\psi_0\rangle$ generates a set of $2^n$ equally
likely states, as considered in Section~\ref{DIST}. In
Fig.~\ref{ptwis03}, the quantities defined in Section~\ref{DIST} are
computed for the twist perturbation~(\ref{twist}) with
perturbation strength $\epsilon=0.03$. Figure~\ref{ptwis03} shows
results after 8 and 12 steps for both the chaotic and the regular
initial conditions.

In the chaotic case in Fig.~\ref{ptwis03}(a), where the total number
of vectors is $2^8=256$ (8 steps), the distribution of Hilbert-space
angles, $g(\phi)$, is concentrated at large angles, i.e., most pairs of
vectors are far apart from each other.  The small peak of $g(\phi)$
at $\phi\simeq\pi/16$ corresponds to 128 pairs of vectors, the two
vectors in each pair being generated by perturbation sequences that
differ only at the first step. The somewhat larger peak of $g(\phi)$
at $\phi\simeq3\pi/16$ similarly indicates the existence of 64 quartets
of vectors, generated by perturbation sequences differing only in the first
two steps. The information $\Delta I$ needed to track a perturbed vector
at resolution level $\phi$ is 8 bits at small angles where each group
contains only one vector. At $\phi\simeq\pi/16$ the information drops
to 7 bits, and at $\phi\simeq3\pi/16$ it drops to 6 bits, reflecting
the grouping of the vectors in pairs and quartets, respectively. For
larger resolution angles, the information stays constant before dropping
rapidly to zero at angles $\phi\agt3\pi/8$. Just as the information
begins to drop rapidly, there is a sudden drop to about 5 bits,
reflecting a further, approximate grouping into 32 octets of vectors,
generated by perturbation sequences that differ only in the first three
steps.  The final drop in the information coincides with the main peak
in the angle distribution $g(\phi)$ and with the rising of the trade-off
entropy to its maximum value of $\Delta H\simeq4$ bits. The number of
explored dimensions is $n_d=\lceil2^{\Delta H_2(\pi/2)}\rceil=19$.

If the number of steps is increased to 12 [see Fig.~\ref{ptwis03}(c)],
the main features of Fig.~\ref{ptwis03}(a) are preserved. The discontinuous
drops in information---from 12 to 11 and from 11 to 10 bits---due to the
formation of pairs and quartets are obvious, but the corresponding
peaks in $g(\phi)$ are now almost invisible due to the larger scale
produced by the larger total number of pairs of vectors. There is now
little evidence of a further grouping into octets.  Indeed,
Figure~\ref{ptwis03}(c) suggests that, apart from the organization
into pairs and quartets, there is not much structure in the distribution
of vectors for a chaotic initial state.  The 1024 quartets seem to
be rather uniformly distributed in a
$n_d=\lceil2^{\Delta H_2(\pi/2)}\rceil=65$-dimensional Hilbert space.
In order to check that the quartets are indeed more or less randomly
distributed in Hilbert space, Fig.~\ref{ptwis03}(c) should be compared
to Fig.~\ref{random}(b), where 1024 random vectors in a 62-dimensional
Hilbert space are shown. The random vectors are chosen at random from
an ensemble distributed uniformly over Hilbert space\cite{Wootters1990}.
The number of dimensions, 62, was chosen by trial and error so that
the total entropy, $\Delta H(\pi/2)$, came out to be the same in
Figures \ref{ptwis03}(c) and \ref{random}(b). Although the angles between
the random vectors are concentrated somewhat more towards larger angles,
there is a striking similarity between these two figures.

As further evidence of the nearly random character of the distribution
in Fig.~\ref{ptwis03}(c), Fig.~\ref{ev} compares the eigenvalues of the
density operators $\hat\rho(\pi/2)$ corresponding to Fig.~\ref{ptwis03}(c)
and to the random vectors in Fig.~\ref{random}(b). The 62 largest
eigenvalues in the chaotic case are almost identical to the 62
eigenvalues corresponding to random vectors in 62-dimensional
Hilbert space.

The distribution of perturbed vectors starting from the regular
initial state is completely different from the chaotic case.
Figure~\ref{ptwis03}(b) shows the regular case after eight steps.
The angle distribution $g(\phi)$ is conspicuously non-random:
it is concentrated at angles smaller than roughly $\pi/4$, and
there is a regular structure of peaks and valleys.  The information
drops rapidly, with little plateaus corresponding to the valleys in
the angle distribution. The number of explored dimensions is
$n_d=2$, which agrees with results of Peres\cite{Peres1991b} that show
that the quantum evolution in a regular region of the kicked top is
essentially confined to a 2-dimensional subspace.

Figure~\ref{ptwis03}(d) shows the regular case after 12 steps. The
average information and the trade-off entropy show a behavior similar
to the 8-step case. The plateaus in the information are washed
out, corresponding to less pronounced minima in the distribution
of angles. This appearance of formerly forbidden angles is expected as
the number of vectors increases; it would occur even for a completely
regular array of vectors. The eigenvalues of the density operator
$\hat\rho(\pi/2)$ corresponding to the 4096 vectors in
Fig.~\ref{ptwis03}(d), shown in Fig.~\ref{ev}, confirm the restriction
to a 2-dimensional subspace. The third-largest eigenvalue, measuring
the relative weight of the third explored dimension, is of order only
$10^{-2}$.

The results shown in Figures \ref{ptwis03} and \ref{ev} display a
striking difference in the distribution of vectors in the chaotic and
regular cases. In the chaotic case, the vectors, aside from the
quartet structure, are distributed randomly in a subspace whose
dimensionality increases with the number of steps. The information
needed to track a perturbed vector after $n$ steps is of the order
of $n$ bits, similar to the information needed to specify a vector
out of a set of $2^n$ random vectors. By contrast, in the
regular case the vectors do not get far apart in Hilbert space,
explore only few dimensions, and do not explore them randomly.

\subsection{Twist perturbation: $\epsilon=0.003$} \label{TWIST2}

Figure~\ref{ptwis003} shows the distribution of vectors arising from
perturbed evolution with a very small twist perturbation of strength
$\epsilon=0.003$. Here, after 12 steps the vectors are not spread very
far in Hilbert space. This is true even in the chaotic case, shown in
Figure~\ref{ptwis003}(a), where a typical angle between vectors is
$\phi=\pi/8$, the information decreases rapidly with the resolution
angle, and only $n_d=6$ dimensions are explored. Even so, the chaotic
case can be easily distinguished from the regular case, shown in
Fig.~\ref{ptwis003}(b), where the perturbation has almost no effect
on the time evolution of the vectors.

To get a picture of the distribution of vectors for a larger number of
steps, Fig.~\ref{ptwis003}(c) shows 4096 vectors selected randomly
from the $2^{200}$ vectors after 200 steps in the chaotic case,
and Fig.~\ref{ptwis003}(d) shows 1024 vectors selected randomly
from the $2^{200}$ vectors after 200 steps in the regular case.
In the chaotic case, the 4096 vectors fill an
$n_d=373$-dimensional subspace quasi-randomly, as can be checked by
comparison with Fig.~\ref{random}(a), where results for 4096 random
vectors in 512 dimensions are shown.  In the regular case, shown in
Fig.~\ref{ptwis003}(d), even after 200 steps not more than $n_d=2$
dimensions are explored. The vectors remain very close together, and
the information drops rapidly with increasing resolution angle. The
difference between the chaotic and regular cases is as striking as
in the previous subsection.

Although the data shown in Fig.~\ref{ptwis003}(c) establish that the
4096 vectors selected from the available $2^{200}$ vectors in the
chaotic case fill a large-dimensional space quasi-randomly, they by no
means establish that the distribution of all $2^{200}$ vectors is similar
to the distribution of $2^{200}$ random vectors.  For example,
Fig.~\ref{ptwis003}(c) would look exactly the same whether the
$2^{200}$ vectors were randomly distributed or were organized into
$2^{200-m}$ randomly distributed groups, each consisting of $2^m$
tightly bunched vectors, provided that the probability to select
more than one vector from any group is negligible---i.e., provided
that $(4096)^2/2^{200-m}=2^{m-176}\ll1$, which is satisfied for
$m\alt170$.  Indeed, in view of the data in the previous subsection,
one expects the vectors to be organized on small angular scales into
pairs and quartets and perhaps into somewhat larger groups that persist
from the first few steps.

To characterize the angle distribution completely would require
the computation of the angles between all pairs among the $2^{200}$
vectors, which would exhaust the storage and computing power of any
computer now and in the foreseeable future.  Our results are thus
rigorous only up to 12 steps, where we are able to compute the angles
between all pairs of vectors.  Nonetheless, our results provide some
support for the conjecture that the distribution in the chaotic case
is essentially random for large numbers of steps.  In order to give
an approximate picture of such a random distribution, we have
developed approximations, shown in Fig.~\ref{ran200}, for the
information and the trade-off entropy for $2^{200}$ random vectors
in a 512-dimensional Hilbert space.

These approximations are based on knowing $N_{D,{\rm max}}(\phi)$,
the maximum number of disjoint spheres of radius $\phi$ that $D$-dimensional
Hilbert space can accommodate.  In Appendix~\ref{AHil} we compute the
volume ${\cal V}_D(\phi)=(\sin\phi)^{2(D-1)}{\cal V}_D$ contained
within a sphere of radius $\phi$ in $D$-dimensional Hilbert space
[Eq.~(\ref{VDPhi})] and the total volume ${\cal V}_D=\pi^{D-1}/(D-1)!$
of $D$-dimensional Hilbert space [Eq.~(\ref{VD})], from which it follows
that
\begin{equation}
N_{D,{\rm max}}(\phi)={{\cal V}_D\over{\cal V}_D(\phi)}
=(\sin\phi)^{-2(D-1)}\;.
\end{equation}
It is worth emphasizing just how enormous Hilbert space is by noting
that the number of spheres of radius $\phi=0.1\,$rad that can be accommodated
within a $D$-dimensional Hilbert space is
$N_{D,{\rm max}}\simeq\phi^{-2(D-1)}=10^{2(D-1)}$; for the 512-dimensional
Hilbert space considered in this paper, this is $10^{1022}$ spheres.

Suppose $N$ vectors, distributed randomly in $D$-dimensional Hilbert
space, are grouped at resolution level $\phi$.  The number of groups
at this resolution, $N(\phi)$, cannot be larger than $N$, the total
number of vectors, nor larger than $N_{D,{\rm max}}(\phi)$, the
maximum number of groups. For the average information, this entails
\begin{equation}
\Delta I(\phi)\leq\log_2N(\phi)\leq\Delta I_{D,{\rm max}}(\phi)
\equiv\min\Bigl(\log_2N_{D,{\rm max}}(\phi),\log_2N\Bigr)\;.
\label{Imax}
\end{equation}
One expects a large number of random vectors to fill Hilbert space
almost uniformly. For angles $\phi$ for which $N_{D,{\rm max}}(\phi)\ll N$,
there are close to $N_{D,{\rm max}}(\phi)$ groups with roughly equal
numbers of vectors in each group; therefore, for those angles
$\phi$, $\Delta I(\phi)\simeq\log_2N_{D,{\rm max}}(\phi)$.  For angles
$\phi$ for which $N_{D,{\rm max}}(\phi)\gg N$, there is just one vector
in each group, whence $\Delta I(\phi)\simeq\log_2N$ for those angles. This
means that the upper bound $\Delta I_{D,{\rm max}}(\phi)$ is an
excellent approximation to $\Delta I(\phi)$ everywhere except for a
small region near the sharp bend located at the angle $\phi_b$ determined
by $N_{D,{\rm max}}(\phi_b)=N$.
The upper bound $\Delta I_{D,{\rm max}}(\phi)$
is plotted in Fig.~\ref{ran200} for $D=512$ and $N=2^{200}$.

The trade-off entropy, on the other hand, cannot be larger than the
maximum entropy of a group, i.e.,
\begin{equation}
\Delta H(\phi)\leq H_{D,{\rm max}}(\phi)\simeq H_D(\phi)\;,
\label{Hmax}
\end{equation}
where $H_{D,{\rm max}}(\phi)$ [Eq.~(\ref{HDmax})] is the maximum
possible entropy for a density operator constructed from vectors that
lie within a sphere of radius $\phi$ in $D$-dimensional Hilbert space,
and $H_D(\phi)$ [Eq.~(\ref{Hsphere})] is the entropy of a uniform
distribution of vectors within a sphere of radius $\phi$.  For
large-dimensional Hilbert spaces, there is no appreciable difference
between $H_{D,{\rm max}}(\phi)$ and $H_D(\phi)$.  The maximum entropy
$H_{D,{\rm max}}(\phi)$ is plotted in Fig.~\ref{ran200} for $D=512$.

For angles $\phi$ for which $N_{D,{\rm max}}(\phi)\ll N$, where the
number of vectors per group is large enough---say, larger than
$D$---the distribution of vectors within each group approximates a
uniform distribution, and thus $\Delta H(\phi)$ is well approximated by
$H_D(\phi)$ and, hence, by $H_{D,{\rm max}}(\phi)$.  This means that
there is a region to the right of $\phi_b$ in Fig.~\ref{ran200}, where
$H_{D,{\rm max}}(\phi)$ is not only an upper bound, but is also a
good approximation to the trade-off entropy.  For angles to the left
of $\phi_b$, the number of vectors in a typical group rapidly
approaches 1, which means that the trade-off entropy is very close
to zero in the region where the average information saturates.

\subsection{Turn perturbation: $\epsilon=0.003$} \label{TURN}

Figure~\ref{prota} displays results for the turn perturbation, showing
the same range of behavior as in the preceding subsections.
Figure~\ref{prota}(a) shows how the $2^{12}$ vectors generated after
12 perturbed steps in the chaotic case fill an $n_d=46$-dimensional
Hilbert space randomly, except for a grouping into pairs, quartets,
and perhaps octets, corresponding to the discontinuous drops in
$\Delta I(\phi)$.  By contrast, in the regular case after 12 steps,
displayed in Fig.~\ref{prota}(b), the vectors
remain close together and fill just $n_d=2$ dimensions.
Figures~\ref{prota}(c) and~(d) show $2^{10}$ vectors chosen randomly
out of the $2^{30}$ vectors generated after 30 perturbed steps.  In the
chaotic case, the distribution can barely be distinguished from the
distribution of random vectors in Fig.~\ref{random}(a). In the
regular case, the vectors are spread a little further apart in
comparison with Fig.~\ref{prota}(b), but they still fill only $n_d=2$
dimensions.

Our results establish well the nearly random character of the distribution
of vectors in the chaotic case. This is the main result of this paper,
providing numerical evidence for hypersensitivity to perturbations in
the quantum kicked top. It is more difficult (and less interesting) to
give a general characterization of the distribution of vectors in the
regular case. One reason for this is the finite size of the regular
region on the classical unit sphere, which makes possible a sort of
diffusion of a perturbed vector out of the regular region into the chaotic
region. Figure~\ref{prota200} investigates this kind of behavior by
showing the distribution of $2^{10}$ vectors randomly chosen after 100
[Fig.~\ref{prota200}(a)] and 200 [Fig.~\ref{prota200}(b)] steps in the
regular case.  It is apparent that the vectors drift more and more
apart and begin to explore more dimensions of Hilbert space, although
even after 200 steps, the number of explored dimensions is still only
$n_d=5$.

Figure~\ref{prota200}(c) shows the eigenvalues of the density operators
obtained by averaging over $2^{10}$ vectors randomly chosen after 30,
100, and 200 perturbed steps in the regular case.  The eigenvalues
provide a more precise picture of the way additional dimensions are
explored, since they are a measure of the relative weight with which the
dimensions of Hilbert space are explored. One sees a slow leaking of
probability into additional dimensions. This leaking is due to the
fact that, with an increasing number of perturbed steps, the
probability increases for a state to have significant support outside
the regular region, i.e., in the chaotic region. The part of the
wave function that is in the chaotic region is subject to chaotic
time evolution and therefore free to explore almost all dimensions of
Hilbert space.

\section{CONNECTION WITH STATISTICAL PHYSICS} \label{STAT}

Consider a physical system, classical or quantum, with a known
Hamiltonian. The {\it state\/} of the system at time $t=t_0$
represents the observer's knowledge of the way the system was
prepared. In classical physics, states are described mathematically by
a probability density $\rho({\bf x})$ in phase space, while in quantum
mechanics, states can be represented either by a Hilbert-space vector
$|\psi\rangle$ or, more generally, by a density operator $\hat\rho$,
depending on the preparation procedure. Observers with different
knowledge assign different states to the system; the state is
therefore not a property of the system alone, but reflects the
observer's state of knowledge about the system.

The entropy (in bits) of a system state, defined in the classical case
as $H=-\int d\Gamma\rho({\bf x})\log_2[\rho({\bf x})]$, where
$d\Gamma$ is the usual phase-space measure, and in the quantum case as
$H=-{\rm Tr}[\hat\rho\log_2(\hat\rho)]$, measures the information missing
toward a complete specification of the system. The classical entropy
is defined up to an arbitrary additive constant, reflecting the fact
that an infinite amount of information would be needed to give the
exact location of a single point in phase space. The quantum entropy
vanishes for a pure state $\hat\rho=|\psi\rangle\langle\psi|$, which is
meaningful because no information beyond that contained in the wave
function exists about a quantum system. As a consequence of
Liouville's theorem, both classical and quantum entropy remain
constant under Hamiltonian time evolution.

To make the connection with thermodynamics, we assume that there is a
heat reservoir at temperature $T_0$, to which all energy in the
form of heat must eventually be transferred, possibly using
intermediate steps such as storage at some lower temperature.
In the presence of this fiducial heat reservoir, the free energy or
maximum average extractable work for an equilibrium state is given
by $F_0=E-T_0k_B\ln2\,H$, where $E$ is the mean internal energy
of the state. (More precisely, it is the difference between the $F_0$
values of two states that determines how much work can be extracted in
a transformation between the two states.)  It is the main premise of
this section that the maximum average extractable work, from now on
called {\it available work}, is given by $F_0=E-T_0k_B\ln2\,H$ {\it
for any state}, even outside equilibrium.  This means that each bit
of missing information costs one $T_0k_B\ln2$ of available work.
General arguments for this premise will be given elsewhere
\cite{Schack1994c}; here we discuss just one important case. If the
non-equilibrium state was formed through reversible Hamiltonian time
evolution starting from an initial equilibrium state, the amount of
work $F_0=E-T_0k_B\ln2\,H$ can {\it in principle\/} be extracted on
average if the system is made to evolve back into the initial
equilibrium state using time reversal. This argument is equivalent
to Loschmidt's famous {\it Umkehreinwand\/}\cite{Tolman1938}.
Although time reversal appears to be impractical in most
situations---a remarkable exception is spin echo---there are no
known fundamental reasons for excluding it.

Since entropy is a measure of the state of knowledge about the system,
and since available work is determined by the entropy, the only way
the available work can change (except for changes in the energy
levels) is via a change in the state of knowledge.  Hamiltonian time
evolution of an isolated system does not lead to a change in the state
of knowledge; entropy and available work remain unchanged. This is a
consequence of Liouville's theorem and is true for regular as well as for
chaotic systems: Hamiltonian time evolution in isolated chaotic
systems does not lead to information loss
\cite{Schack1993e,Caves1993b,Schack1992a}. The following paragraphs
discuss the three ways in which information about the system can
change: measurement, deliberate discarding of information, and
interaction with an incompletely known environment.

Available work can {\it increase\/} if an observation is made on the
system. The accompanying decrease in entropy does not constitute a
violation of the second law of thermodynamics, however, because the
physical state of the observer changes in the course of the
observation.  Landauer\cite{Landauer1961}, following seminal work by
Szilard\cite{Szilard1929}, has provided a simple and elegant
quantitative description of the change in the state of the observer.
If the observer wants to use additional information about the system
to increase the available work, he must keep a physical record of the
information. According to Landauer's principle, the erasure of a bit
of information in the presence of a reservoir at temperature $T_0$ is
necessarily accompanied by dissipation of at least an amount
$T_0k_B\ln2$ of energy. If this thermodynamic cost of erasing information
is taken into account as a negative contribution to free energy, no
observation can increase the total available work on the average
\cite{Zurek1989a,Zurek1989b,Caves1990c}.  Information that the observer
possesses about the system therefore plays a role complementary to the
entropy or missing information. Entropy or missing information about
the system reduces the available work through the usual entropy term
in the free energy; information the observer actually possesses must
be stored physically, thus reducing further the available work due to
the Landauer erasure cost. {\it Total\/} available work is determined
by the sum of entropy and
information\cite{Zurek1989a,Zurek1989b,Caves1990c,Caves1993b}.

Available work can {\it decrease\/} if information about the system is
lost. Information loss is equivalent to entropy increase. There are
two main mechanisms leading to information loss: deliberate discarding
of information and loss of information through interaction with an
incompletely known environment. It must be emphasized that the
well-known sensitivity to initial conditions in classical chaotic
systems does not entail information loss, because statistical physics
is concerned with the time evolution of probability distributions
governed by Liouville's equation, not with trajectories of single
phase-space points.

Deliberate discarding of information was used by
Jaynes\cite{Jaynes1957a,Jaynes1957b,Jaynes1983} to derive traditional
thermodynamics. Jaynes showed how equilibrium thermodynamics follows
effortlessly from Liouville's equation if only information about the
values of the macroscopic variables defining a thermodynamic state is
retained. In Jaynes's approach, irrelevant information is discarded by
means of the principle of maximum entropy. Another way to discard
information considered irrelevant is coarse graining, where
all details of a state below a certain scale are ignored.

In contrast to these examples where information is discarded
deliberately, an actual loss of information can occur in a system
that, rather than being perfectly isolated, interacts with an
incompletely known environment. The interaction with the environment
leads to a perturbed time evolution of the system. Predictions for the
system state are made by tracing out the environment, i.e., by
averaging over the perturbations, which is generally accompanied by an
entropy increase.

Nothing forces one, however, to average over the perturbations.
Alternatively one could, by making observations on the environment,
gather enough information about the perturbations to keep track of the
perturbed evolved system state to a certain accuracy, thereby reducing
the entropy increase. In Section~\ref{DIST}, with the accuracy determined
by the resolution angle $\phi$, the minimum information needed to keep
track of the system state to accuracy $\phi$ was denoted by $\Delta
I(\phi)$, and the resulting average entropy increase, the trade-off
entropy, was denoted by $\Delta H(\phi)$. Averaging over the
perturbations corresponds to an accuracy $\phi=\pi/2$: $\Delta
I(\pi/2)$ vanishes, and $\Delta H(\pi/2)$ is the entropy due to
averaging over the perturbation. Equation~(\ref{Balian}) shows that
the sum of information and trade-off entropy is never less than the
entropy due to averaging, so that one can never gain in terms of total
available work by gathering information about the perturbations. But
at this stage, one has no reason to expect that one would do much
worse by keeping track of the system state, so that in principle, the
system entropy could be kept from increasing.

For a system showing hypersensitivity to perturbation, however, there
is a compelling reason not to keep track of a perturbed system state,
but to average over the perturbations. The information needed to keep
track of a perturbed state increases far more rapidly than the entropy
due to averaging, which means that keeping track of the perturbed
state would lead to an enormous reduction in available work due to the
thermodynamic erasure cost. We have
conjectured\cite{Schack1993e,Caves1993b,Schack1992a} that
hypersensitivity to perturbation provides a quantitative link between
chaos and entropy increase in both classical and quantum open systems.
Within the limits of our numerical method, the present paper
establishes this link for the quantum kicked top.

\acknowledgements
One of us (RS) acknowledges the support of a fellowship from the Deutsche
Forschungsgemeinschaft.

\appendix

\section{Volume contained within a sphere in Hilbert space}
\label{AHil}

In this Appendix we compute the volume contained within a sphere
of radius $\Phi$ in $D$-dimensional Hilbert space.  More precisely,
we work in projective Hilbert space, i.e., the space of Hilbert-space
rays or the space of normalized state vectors in which vectors that
differ by a phase factor are equivalent.

We begin by deriving the line element of the Riemannian metric
that corresponds to the Hilbert-space angle~(\ref{Hdistance}).
Consider two neighboring normalized state vectors $|\psi\rangle$
and $|\psi\rangle+|d\psi\rangle$.  The infinitesimal angle $ds$
between these vectors satisfies
\begin{eqnarray}
1-ds^2=\cos^2ds
&=&\Bigl|\Bigl\langle\psi\Bigm|
(|\psi\rangle+|d\psi\rangle)\Bigr.\Bigr|^2\nonumber\\
&=&1+2{\rm Re}(\langle\psi|d\psi\rangle)+|\langle\psi|d\psi\rangle|^2\;.
\end{eqnarray}
The normalization of $|\psi\rangle$ and $|\psi\rangle+|d\psi\rangle$
implies that
\begin{equation}
2{\rm Re}(\langle\psi|d\psi\rangle)=-\langle d\psi|d\psi\rangle\;,
\label{norm1}
\end{equation}
so the line element becomes
\begin{equation}
ds^2=\langle d\psi|d\psi\rangle-|\langle\psi|d\psi\rangle|^2
=\langle d\psi_\perp|d\psi_\perp\rangle\;,
\label{FSmetric}
\end{equation}
where
$|d\psi_\perp\rangle=|d\psi\rangle-|\psi\rangle\langle\psi|d\psi\rangle$
is the projection of $|d\psi\rangle$ orthogonal to $|\psi\rangle$.
The metric~(\ref{FSmetric}), called the Fubini-Study metric
\cite{Gibbons1992}, is the natural metric on projective Hilbert space.
Notice that the line element is invariant under phase changes of
$|\psi\rangle$ and $|\psi\rangle+|d\psi\rangle$.

Consider now a sphere of radius $\Phi\le\pi/2$ in projective Hilbert
space; let the center of the sphere be denoted by $|\psi_0\rangle$.
Any normalized vector $|\psi\rangle$ can be written as
\begin{equation}
|\psi\rangle=
e^{i\delta}\cos\phi|\psi_0\rangle+\sin\phi|\eta\rangle\;,
\end{equation}
where $|\eta\rangle$ is a normalized vector in the subspace orthogonal
to $|\psi_0\rangle$, and the polar angle
$\phi=\cos^{-1}(|\langle\psi|\psi_0\rangle|)$ satisfies $0\le\phi\le\pi/2$.
The region contained within our sphere of radius $\Phi$, which we denote by
$V_D(\Phi)$, consists of all vectors such that $\phi\le\Phi$.  The
phase freedom in $|\psi\rangle$ can be removed by choosing the phase
$\delta=0$.  That having been done, $|\eta\rangle$ ranges over all
normalized vectors in the subspace orthogonal to $|\psi_0\rangle$; in
particular, two normalized vectors $|\eta\rangle$, differing only by
a phase factor, are {\it not\/} equivalent.

We can now write
\begin{equation}
|d\psi\rangle=-\sin\phi\,d\phi|\psi_0\rangle+
\cos\phi\,d\phi|\eta\rangle+\sin\phi|d\eta\rangle\;,
\end{equation}
where $|d\eta\rangle$ is the infinitesimal change in $|\eta\rangle$
(notice that $\langle\psi_0|d\eta\rangle=0$).  Normalization of
$|\eta\rangle$ and $|\eta\rangle+|d\eta\rangle$ implies, just as in
Eq.~(\ref{norm1}), that
\begin{equation}
2{\rm Re}(\langle\eta|d\eta\rangle)=-\langle d\eta|d\eta\rangle\;.
\label{norm2}
\end{equation}
Retaining only second-order terms in the infinitesimal changes, we can
put the line element~(\ref{FSmetric}) in the form
\begin{equation}
ds^2=d\phi^2+\sin^2\phi\,d\gamma^2\;,
\label{FSmetric2}
\end{equation}
where
\begin{equation}
d\gamma^2=
\langle d\eta|d\eta\rangle
-\sin^2\phi\,|\langle\eta|d\eta\rangle|^2
\label{modsphere}
\end{equation}
defines a Riemannian metric on the the space of normalized vectors in
the ($D-1$)-dimensional subspace orthogonal to $|\psi_0\rangle$.  This
space is a ($2D-3$)-dimensional sphere of unit radius, denoted
$S_{2D-3}$, although the line element $d\gamma^2$ is, as we show
below, different from the standard geometry on a ($2D-3$)-dimensional
unit sphere.  Notice that the Fubini-Study metric~(\ref{FSmetric2})
scales all lengths on $S_{2D-3}$ by a factor of $\sin\phi$; this scaling
is analogous to the way that polar angle on an ordinary 2-sphere scales
the size of circles (1-spheres) of latitude.

Consider now any orthonormal basis $|\eta_j\rangle$, $j=1,\ldots,D-1$,
in the subspace orthogonal to $|\psi_0\rangle$.  We can introduce
coordinates on  $S_{2D-3}$ by expanding
$|\eta\rangle$ as
\begin{equation}
|\eta\rangle=\sum_{j=1}^{D-1}(x_j+iy_j)|\eta_j\rangle\;.
\end{equation}
Normalization of $|\eta\rangle$ implies the constraint
\begin{equation}
1=\sum_{j=1}^{D-1}x_j^2+y_j^2\;,
\label{constraint}
\end{equation}
which defines the ($2D-3$)-dimensional unit sphere.  The first term
in the metric~(\ref{modsphere}),
\begin{equation}
\langle d\eta|d\eta\rangle=\sum_{j=1}^{D-1}dx_j^2+dy_j^2\;,
\end{equation}
is the flat Euclidean metric in $2(D-1)$ dimensions; it induces the
standard metric on  $S_{2D-3}$.  The second term in the
metric~(\ref{modsphere}) modifies the standard geometry on  $S_{2D-3}$.
Noting that
\begin{equation}
\langle\eta|d\eta\rangle=
{1\over2}\,d\!\left(\sum_{j=1}^{D-1}x_j^2+y_j^2\right)
+i\sum_{j=1}^{D-1}x_jdy_j-y_jdx_j=
i\sum_{j=1}^{D-1}x_jdy_j-y_jdx_j\;,
\end{equation}
we can put the second term in~(\ref{modsphere}) in the form
\begin{equation}
\sin^2\phi\,|\langle\eta|d\eta\rangle|^2=
\sin^2\phi\left(\sum_{j=1}^{D-1}x_jdy_j-y_jdx_j\right)^2\;.
\end{equation}

Perhaps the easiest way to see how the second term affects the geometry
on  $S_{2D-3}$ is to make a judicious choice of coordinates.  Given an
arbitrary vector $|\eta\rangle$, we can always choose the orthonormal
basis so that $|\eta_1\rangle=|\eta\rangle$, which means that
$|\eta\rangle$ is assigned coordinates $x_1=1$, $y_1=0$, and
$x_j=y_j=0$, $j=2,\ldots,D-1$.  In these special coordinates the
metric~(\ref{modsphere}), evaluated at $|\eta\rangle$, takes the form
\begin{equation}
d\gamma^2=\cos^2\phi\,dy_1^2+\sum_{j=2}^{D-1}dx_j^2+dy_j^2\;,
\end{equation}
where we have used the constraint~(\ref{constraint}) to write
\begin{equation}
dx_1={1\over x_1}\left(-y_1dy_1-\sum_{j=2}^{D-1}x_jdx_j+y_jdy_j\right)
=0\;\;\mbox{at $|\eta\rangle$.}
\end{equation}
In these special coordinates the standard geometry on  $S_{2D-3}$
has the same form at $|\eta\rangle$, except that the $\cos\phi$ is
replaced by 1.  The effect of the second term in the
metric~(\ref{modsphere}) is thus to shorten lengths (relative to the
standard geometry) along one direction on  $S_{2D-3}$ by a factor
$\cos\phi$; the direction of shortened length corresponds, in the
special coordinates, to the $y_1$ direction or, in coordinate-free
language, to an infinitesimal change in the phase of $|\eta\rangle$.

In the special coordinates the volume element on $S_{2D-3}$ at
$|\eta\rangle$, defined by the line element $d\gamma^2$, is given by
\begin{equation}
\cos\phi\,dy_1dx_2\ldots dx_{D-1}dy_2\ldots dy_{D-1}
=\cos\phi\,d{\cal S}_{2D-3}\;,
\label{gammavolume}
\end{equation}
where $d{\cal S}_{2D-3}$ is the volume element defined by the standard
geometry on  $S_{2D-3}$.  Writing the volume element~(\ref{gammavolume})
in terms of $d{\cal S}_{2D-3}$ frees it from dependence on the special
coordinates.  Referring to the Fubini-Study metric~(\ref{FSmetric2}),
we can now write the volume element on projective Hilbert space as
\begin{equation}
d{\cal V}_D=(\sin\phi)^{2D-3}\cos\phi\,d\phi\,d{\cal S}_{2D-3}\;.
\label{dVD}
\end{equation}
The $2D-3$ factors of $\sin\phi$ come from scaling all $2D-3$ dimensions
of $S_{2D-3}$, as required by the Fubini-Study metric~(\ref{FSmetric2}).

We are now prepared to compute the volume ${\cal V}_D(\Phi)$ contained
within a Hilbert-space sphere of radius $\Phi$:
\begin{eqnarray}
{\cal V}_D(\Phi)=\int_{V_D(\Phi)}d{\cal V}_D
&=&\int_0^\Phi d\phi\,(\sin\phi)^{2D-3}\cos\phi\int d{\cal S}_{2D-3}\nonumber\\
&=&{{\cal S}_{2D-3}\over2(D-1)}(\sin\Phi)^{2(D-1)}
=(\sin\Phi)^{2(D-1)}{\cal V}_D\;.
\label{VDPhi}
\end{eqnarray}
Here ${\cal S}_{2D-3}$ is the volume of the unit sphere $S_{2D-3}$,
calculated using the standard geometry (Be careful: this is the
``area'' of $S_{2D-3}$, not the volume interior to it), and
\begin{equation}
{\cal V}_D={\cal V}_D(\pi/2)={{\cal S}_{2D-3}\over2(D-1)}
\end{equation}
is the total volume of projective Hilbert space.

The volume of an $n$-dimensional unit sphere $S_n$,
\begin{equation}
{\cal S}_n=
{2\pi^{(n+1)/2}\over\displaystyle{\Gamma\!\left({n+1\over2}\right)}}\;,
\end{equation}
follows from a standard trick involving Gaussian integrals:
\begin{eqnarray}
\case1/2 S_n\,\Gamma\!\left({n+1\over2}\right)&=&
\case1/2 S_n\int_0^\infty dv\,v^{(n-1)/2}e^{-v}\nonumber\\
&=&\int_0^\infty dr\,r^nS_ne^{-r^2}=
\left(\int_{-\infty}^\infty du\,e^{-u^2}\right)^{n+1}=\pi^{(n+1)/2}\;.
\end{eqnarray}
This gives us ${\cal S}_{2D-3}=2\pi^{D-1}/(D-2)!$, which allows us to
write the total volume of projective Hilbert space \cite{Gibbons1992} as
\begin{equation}
{\cal V}_D={\pi^{D-1}\over(D-1)!}\;.
\label{VD}
\end{equation}

We can obtain immediately two other results: (i)~the volume contained
between two Hilbert-space spheres of radius $\phi$ and $\phi+d\phi$ is
\begin{equation}
d{\cal V}_D(\phi)={\cal S}_{2D-3}(\sin\phi)^{2D-3}\cos\phi\,d\phi=
{\cal V}_D2(D-1)(\sin\phi)^{2D-3}\cos\phi\,d\phi\;;
\end{equation}
(ii)~the probability that two vectors selected at random are separated by
a Hilbert-space angle between $\phi$ and $\phi+d\phi$ is
\begin{equation}
g(\phi)d\phi={d{\cal V}_D(\phi)\over{\cal V}_D}=
2(D-1)(\sin\phi)^{2D-3}\cos\phi\,d\phi\;.
\label{gphi}
\end{equation}

\section{Entropies of distributions of vectors within a Hilbert-space sphere}
\label{AEnt}

In this Appendix we compute entropies of density operators constructed
from vectors that lie within a sphere of radius $\Phi$ in $D$-dimensional
Hilbert space.  We first compute the entropy $H_D(\Phi)$ of a density
operator $\hat\rho$ that corresponds to a uniform distribution of
Hilbert-space vectors in the region $V_D(\Phi)$ contained within a
sphere of radius $\Phi$.  Formally, $\hat\rho$ is given by
\begin{equation}
\hat\rho=\int_{V_D(\Phi)}\frac{d{\cal V}_D}{{\cal V}_D(\Phi)}\,
|\psi\rangle\langle\psi|\;,
\end{equation}
where ${\cal V}_D(\Phi)=(\sin\Phi)^{2(D-1)}{\cal V}_D$ [Eq.~(\ref{VDPhi})]
is the volume contained within a sphere of radius $\Phi$ in $D$-dimensional
Hilbert space, and
$d{\cal V}_D=(\sin\phi)^{2D-3}\cos\phi\,d\phi\,d{\cal S}_{2D-3}$
[Eq.~(\ref{dVD})] is the volume element on projective Hilbert
space.  Let the center of the sphere be denoted by $|\psi_0\rangle$.
Symmetry about $|\psi_0\rangle$ entails that $\hat\rho$ have the form
\begin{equation}
\hat\rho=\lambda_0\,|\psi_0\rangle\langle\psi_0|+
\frac{1-\lambda_0}{D-1}\,(\hat1-|\psi_0\rangle\langle\psi_0|)\;.
\label{symmetricdo}
\end{equation}
The eigenvalues of $\hat\rho$ are
\begin{eqnarray}
\lambda_0&=&\langle\psi_0|\hat\rho|\psi_0\rangle =
\int_{V_D(\Phi)}\frac{d{\cal V}_D}{{\cal V}_D(\Phi)}\,
|\langle\psi_0|\psi\rangle|^2
=\int_{V_D(\Phi)}\frac{d{\cal V}_D}{{\cal V}_D(\Phi)}\,
\cos^2\phi \nonumber  \\
&=&\frac{{\cal S}_{2D-3}}{(\sin\Phi)^{2(D-1)}{\cal V}_D}
         \int_0^\Phi d\phi\,(\sin\phi)^{2D-3}\cos^3\phi \nonumber \\
&=&1-\frac{D-1}{D}\sin^2\Phi\ge{1\over D}
\end{eqnarray}
and
\begin{equation}
\lambda_k=\frac{1-\lambda_0}{D-1}=\frac{\sin^2\Phi}{D}\le{1\over D}\;,\;\;
k=1,\ldots,D-1\;.
\end{equation}

The entropy of $\hat\rho$ is thus
\begin{eqnarray}
H_D(\Phi)&=&-\sum_{k=0}^{D-1}\lambda_k\log_2\lambda_k\nonumber\\
&=&-\lambda_0\log_2\lambda_0-(1-\lambda_0)\log_2(1-\lambda_0)
+(1-\lambda_0)\log_2(D-1)
\label{Hsymmetricdo}\\
&=&\left(\frac{D-1}{D}\sin^2\Phi-1\right)
\log_2\!\left(1-\frac{D-1}{D}\sin^2\Phi\right)\nonumber\\
&\mbox{}&\;\;-\frac{D-1}{D}\sin^2\Phi\,
\log_2\!\left(\frac{\sin^2\Phi}{D}\right)\;.
\label{Hsphere}
\end{eqnarray}
For comparison, the spread of $\hat\rho$ [Eq.~(\ref{defspread})] has
its maximum value, regardless of the value of $\Phi$,
\begin{equation}
H_{2,D}(\Phi)=-\sum_{k=1}^{D-1}{\lambda_k\over1-\lambda_0}
\log_2\!\left({\lambda_k\over1-\lambda_0}\right)=\log_2(D-1)\;,
\end{equation}
reflecting the fact that the uniform distribution of vectors within
a sphere of radius $\Phi$ explores all $D-1$ dimensions in the
subspace orthogonal to $|\psi_0\rangle$.  This result regarding
the spread does not depend on the particular value of $\lambda_0$:
any density operator of the symmetric form~(\ref{symmetricdo}) has
the maximum spread, $\log_2(D-1)$, provided only that $\lambda_0$
is the largest eigenvalue, i.e., $\lambda_0\ge1/D$
[cf.~Eqs.~(\ref{entropyspread}) and (\ref{Hsymmetricdo})].

It is interesting to compare the entropy of a uniform distribution
of vectors in $V_D(\Phi)$ with the maximum entropy that can be attained
by distributing vectors in $V_D(\Phi)$.  To do this, consider a density
operator
\begin{equation}
\hat\rho=\sum_{i=1}^N p_i|\psi_i\rangle\langle\psi_i|\;,
\end{equation}
constructed from vectors $|\psi_i\rangle$ that lie in $V_D(\Phi)$, i.e.,
$|\langle\psi_0|\psi_i\rangle|=\cos\phi_i\ge\cos\Phi$, $i=1,\ldots,N$.

A unitary transformation in the subspace orthogonal to $|\psi_0\rangle$
leaves the angles $\phi_i$ and the entropy unchanged.  Mixing the density
operators that result from all such unitary transformations gives a new
density operator which is symmetric about $|\psi_0\rangle$, but which,
because of the concavity of the entropy \cite{Balian1991,Caves1994a}, has
an entropy that is not smaller than the entropy of $\hat\rho$.  Thus, in
seeking the maximum entropy, we can restrict attention to density
operators that are symmetric about $|\psi_0\rangle$ and, hence, have the
form~(\ref{symmetricdo}) and have entropy given by Eq.~(\ref{Hsymmetricdo}).
This entropy, which is a function of the eigenvalue $\lambda_0$, has
a maximum value of $\log_2 D$ at $\lambda_0=1/D$ and decreases
monotonically away from this maximum in either direction.

The eigenvalue $\lambda_0$ is bounded below by
\begin{equation}
\lambda_0=\langle\psi_0|\hat\rho|\psi_0\rangle=
\sum_{i=1}^N p_i\cos^2\phi_i\ge\cos^2\Phi\;.
\end{equation}
Hence, an upper bound on the entropy follows from choosing
$\lambda_0=\cos^2\Phi$ when $\cos^2\Phi\ge1/D$ and choosing
$\lambda_0=1/D$ when $\cos^2\Phi\le1/D$.  The upper bound
is given by
\begin{eqnarray}
&\mbox{}&H_{D,{\rm max}}(\Phi)\nonumber\\
&\mbox{}&\;\mbox{}=\left\{\begin{array}{ll}
	   -\cos^2\Phi\,\log_2\cos^2\Phi-\sin^2\Phi\,\log_2\sin^2\Phi
	   +\sin^2\Phi\,\log_2(D-1)\;,\;&\cos^2\Phi\ge1/D\;,\\
	   \log_2 D\;,&\cos^2\Phi\le1/D\;.
              	          \end{array}
                   \right.
\label{HDmax}
\end{eqnarray}
This function is plotted in Fig.~\ref{ran200} for $D=512$.  Notice that
the entropy~(\ref{Hsphere}) of a uniform distribution of vectors within
a sphere of radius $\Phi$ approaches the upper bound~(\ref{HDmax}) as
$D\rightarrow\infty$.

That the bound is actually the maximum, as implied by the notation, is
demonstrated by finding a density operator, constructed from vectors
in $V_D(\Phi)$, which achieves the entropy upper bound.  To that end,
consider the $2(D-1)$ vectors
\begin{eqnarray}
|\psi_j\rangle&=&\cos\Phi|\psi_0\rangle+\sin\Phi|\eta_j\rangle\;,\;\;
j=1,\ldots,D-1\;,\nonumber\\
|\psi'_j\rangle&=&\cos\Phi|\psi_0\rangle-\sin\Phi|\eta_j\rangle\;,\;\;
j=1,\ldots,D-1\;,
\label{symvectors}
\end{eqnarray}
where the vectors $|\eta_j\rangle$ make up an orthonormal basis in the
subspace orthogonal to $|\psi_0\rangle$.  The vectors~(\ref{symvectors})
all lie on the sphere of radius $\Phi$---as far from $|\psi_0\rangle$ as
is allowed.  Construct the density operator
\begin{eqnarray}
\hat\rho&=&p_0|\psi_0\rangle\langle\psi_0|+
{1-p_0\over2(D-1)}\sum_{j=1}^{D-1}
|\psi_j\rangle\langle\psi_j|+|\psi'_j\rangle\langle\psi'_j|\nonumber\\
&=&(\cos^2\Phi+p_0\sin^2\Phi)|\psi_0\rangle\langle\psi_0|+
{1-p_0\over (D-1)}\sin^2\Phi\,(\hat 1-|\psi_0\rangle\langle\psi_0|)\;,
\label{maxdo}
\end{eqnarray}
which has the symmetric form of Eq.~(\ref{symmetricdo}), with
$\lambda_0=\cos^2\Phi+p_0\sin^2\Phi$.  One would obtain the
density operator~(\ref{maxdo}) by letting
$\hat\rho-p_0|\psi_0\rangle\langle\psi_0|$ be constructed from any
set of vectors that lie on the sphere of radius $\Phi$ and are
symmetrically distributed about $|\psi_0\rangle$.  To achieve the upper
bound~(\ref{HDmax}), one chooses $p_0=0$ when $\cos^2\Phi\ge1/D$ and
chooses $p_0=(1/D-\cos^2\Phi)/\sin^2\Phi$ when $\cos^2\Phi\le1/D$.



\begin{figure}
\caption{Distribution of $2^n$ vectors randomly chosen in
$D$-dimensional
Hilbert space. Each diagram shows, as a function of the angle $\phi$,
the distribution $g(\phi)$ of Hilbert-space angles (unnormalized, in
arbitrary units), the average information $\Delta I(\phi)$ to specify
a vector at the resolution given by $\phi$ (in bits), the trade-off
entropy $\Delta H(\phi)$ (in bits), and the average spread $\Delta
H_2(\phi)$ (in bits). For a precise definition of these quantities,
see Section~\protect\ref{DIST}. (a) $n=12$, $D=512$.  (b) $n=10$,
$D=62$.}
\label{random}
\end{figure}

\begin{figure}
\caption{Results characterizing the distribution of Hilbert-space
vectors for the perturbed kicked top with $J=511.5$ and $k=3$
in the presence of
the twist perturbation (\protect\ref{twist}) with $\epsilon=0.03$.
The same quantities as in Fig.~\protect\ref{random} are shown. (a)
Chaotic case, i.e., initial coherent state $|\theta,\varphi\rangle$
centered in the chaotic region with $\theta$ and $\varphi$ given by
Eq.~(\protect\ref{hyperbolic}).  Distribution of all $2^8$ vectors
generated after 8 perturbed steps.  (b) Regular case, i.e., initial
coherent state centered at the elliptic fixed point given by
Eq.~(\protect\ref{elliptic}). All $2^8$ vectors generated after 8
perturbed steps. (c) Chaotic case. 12 steps, all $2^{12}$ vectors. (d)
Regular case. 12 steps, all $2^{12}$ vectors.}
\label{ptwis03}
\end{figure}

\begin{figure}
\caption{Eigenvalues of the density operators formed by averaging over
random vectors (squares), vectors generated by the perturbed kicked
top in the chaotic case (diamonds), and vectors generated by the
perturbed kicked top in the regular case (crosses). The density
operator in the random case was generated by averaging over 1024
vectors randomly chosen in 62-dimensional Hilbert space. The density
operators in the chaotic and regular cases were formed from the
$2^{12}$ vectors in Figs.~\protect\ref{ptwis03}(c) and (d),
respectively. The main diagram shows, for all three cases, the largest
62 eigenvalues greater than $10^{-10}$. The inset shows all
eigenvalues greater than $10^{-20}$ for the chaotic case (solid line)
and the regular case (dashed line).}
\label{ev}
\end{figure}

\begin{figure}
\caption{As Fig.~\protect\ref{ptwis03}, but using the twist
perturbation (\protect\ref{twist}) with $\epsilon=0.003$. (a) Chaotic
case. 12 steps, all $2^{12}$ vectors. (b) Regular case. 12 steps, all
$2^{12}$ vectors. (c) Chaotic case. $2^{12}$ vectors randomly chosen
after 200 perturbed steps. (d) Regular case. $2^{10}$ vectors randomly
chosen after 200 perturbed steps.}
\label{ptwis003}
\end{figure}

\begin{figure}
\caption{Upper bounds for the average information,
$\Delta I_{D,{\rm max}}$, and for the trade-off entropy,
$H_{D,{\rm max}}$, for $2^{200}$ random vectors in 512-dimensional
Hilbert space, as given by Eqs.~(\protect\ref{Imax})
and~(\protect\ref{HDmax}). The curve for $\Delta I_{D,{\rm max}}$ is
an excellent approximation to the average information $\Delta I(\phi)$
except for a small region around $\phi_b$.  The trade-off entropy
$\Delta H(\phi)$ is well approximated by the curve for
$H_{D,{\rm max}}(\phi)$ for angles above $\phi_b$,
but goes to zero rapidly for angles below $\phi_b$.}
\label{ran200}
\end{figure}

\begin{figure}
\caption{As Fig.~\protect\ref{ptwis03}, but using the turn
perturbation (\protect\ref{turn}) with $\epsilon=0.003$. (a) Chaotic
case. 12 steps, all $2^{12}$ vectors. (b) Regular case. 12 steps, all
$2^{12}$ vectors. (c) Chaotic case. $2^{10}$ vectors randomly chosen
after 30 perturbed steps. (d) Regular case. $2^{10}$ vectors randomly
chosen after 30 perturbed steps.}
\label{prota}
\end{figure}

\begin{figure}
\caption{As Fig.~\protect\ref{prota}, regular case. (a) $2^{10}$ vectors
randomly chosen after 100 perturbed steps. (b) $2^{10}$ vectors
randomly chosen after 200 perturbed steps. (c) The 20 largest
eigenvalues of the density operators obtained by averaging over
$2^{10}$ vectors randomly chosen after 30, 100, and 200 perturbed
steps.}
\label{prota200}
\end{figure}

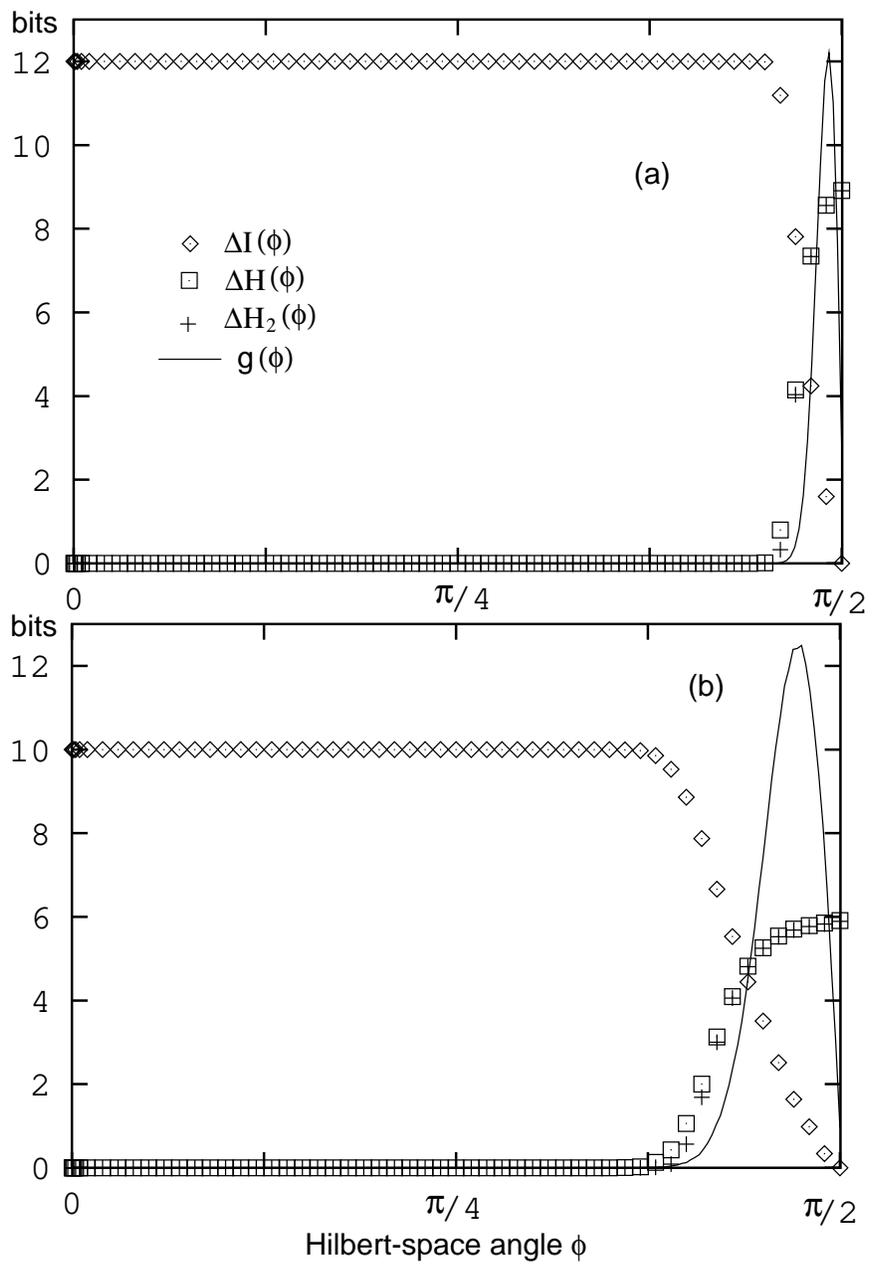
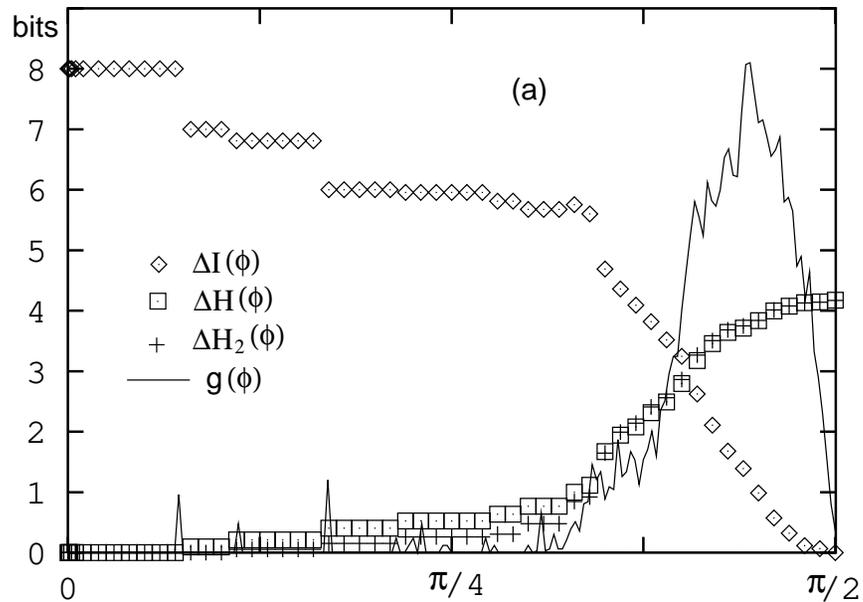
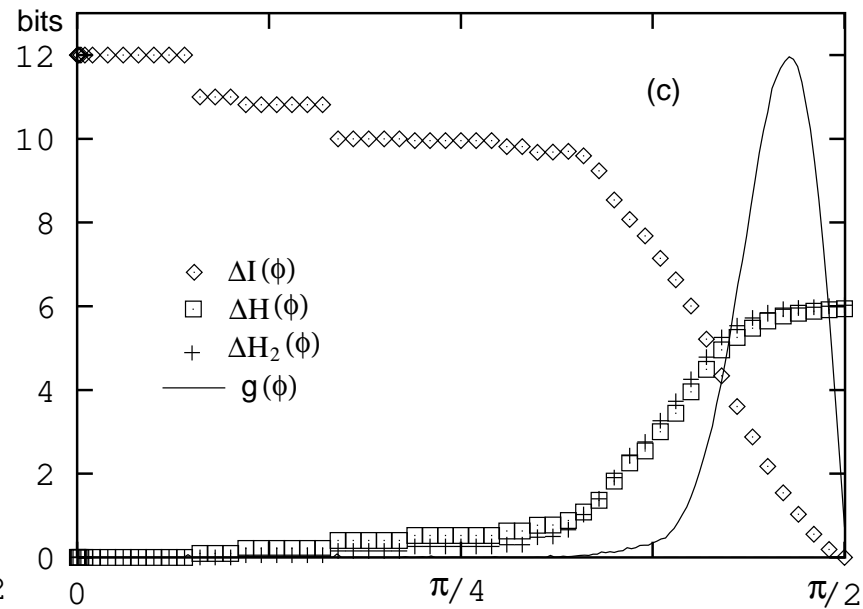
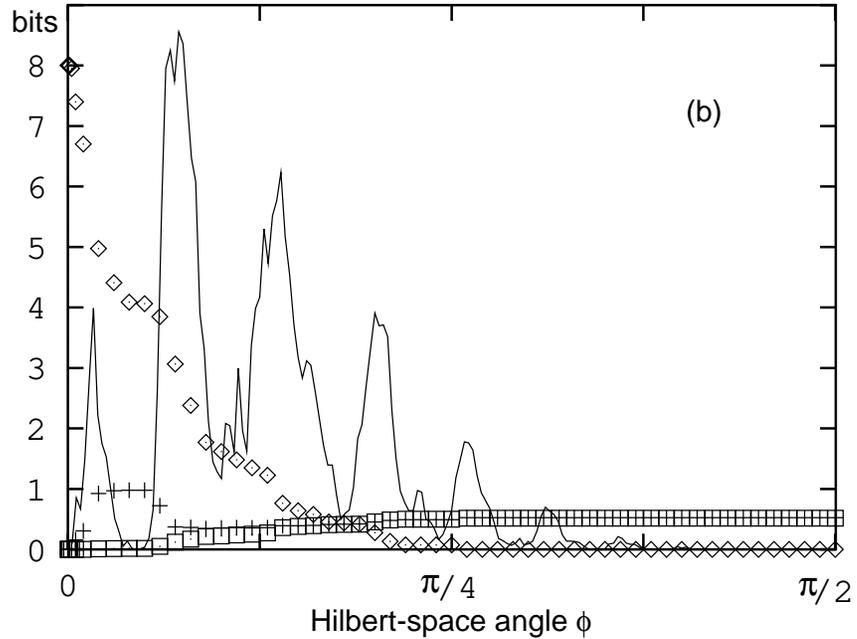
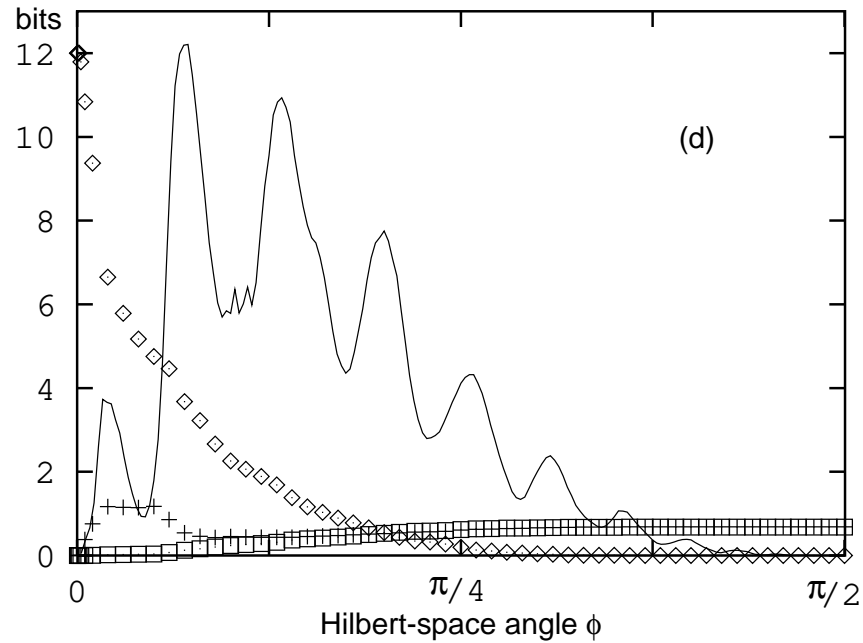
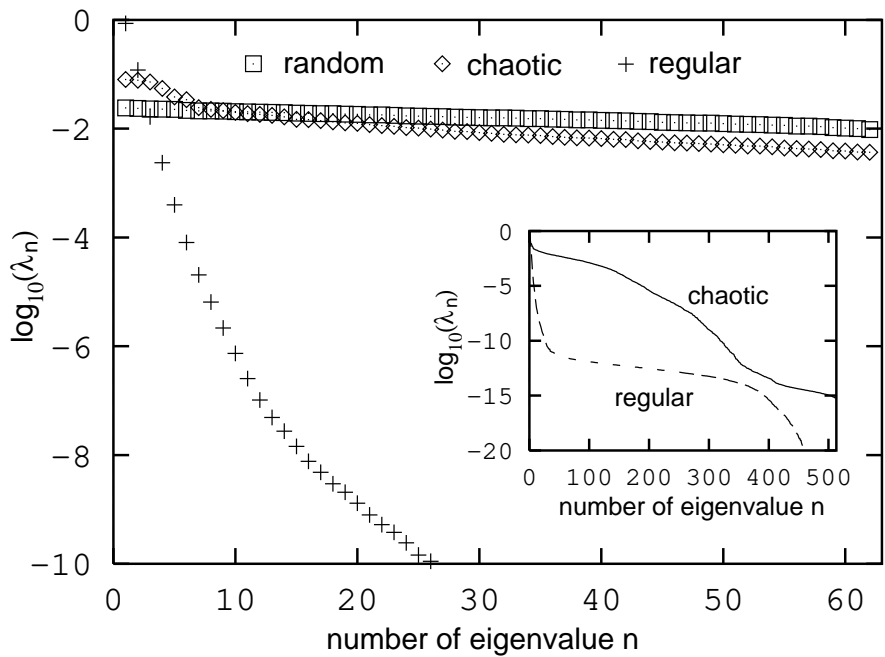
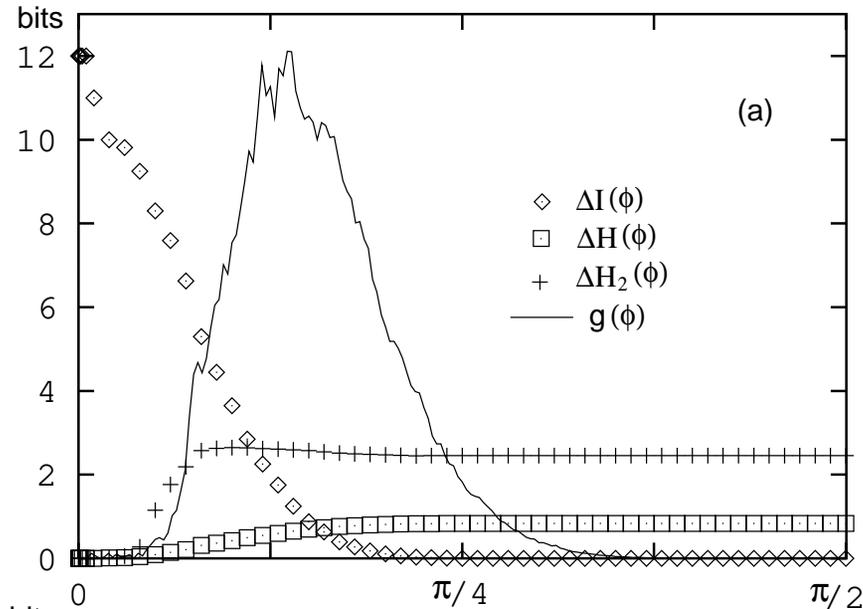
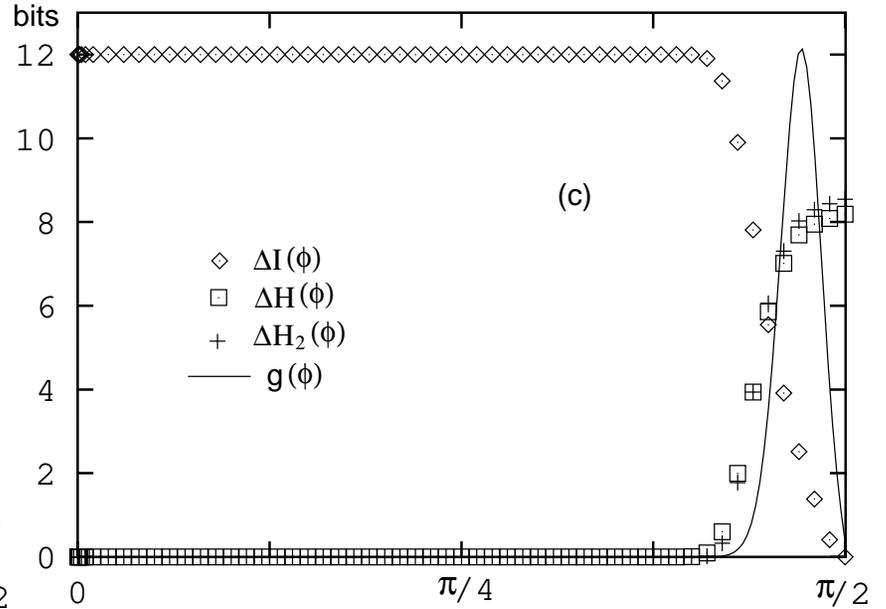
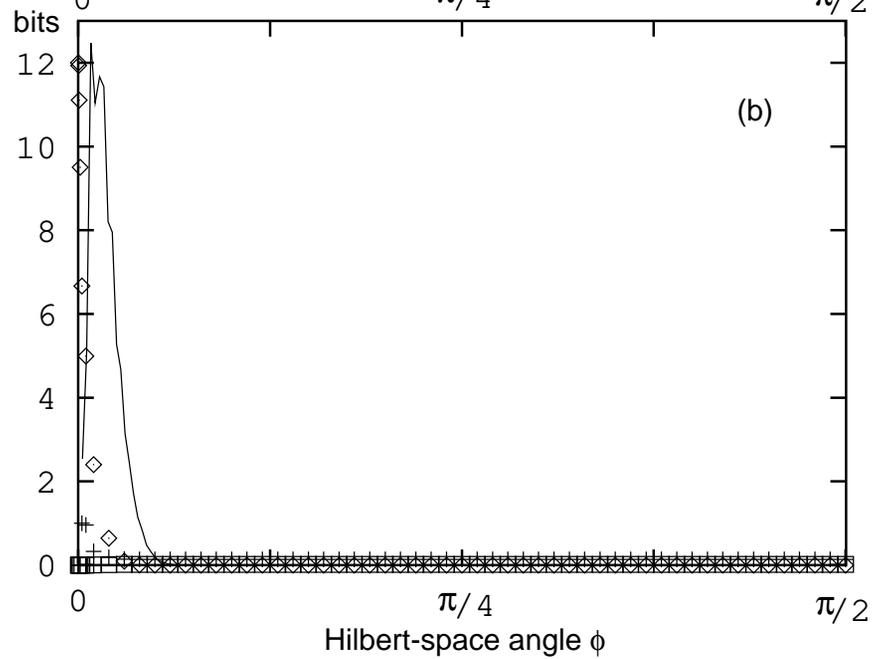
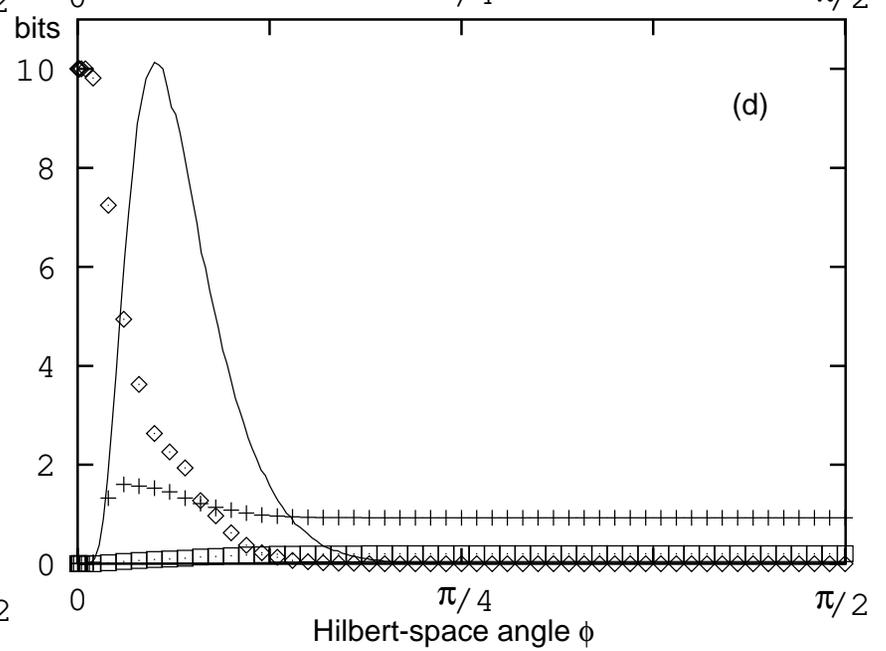
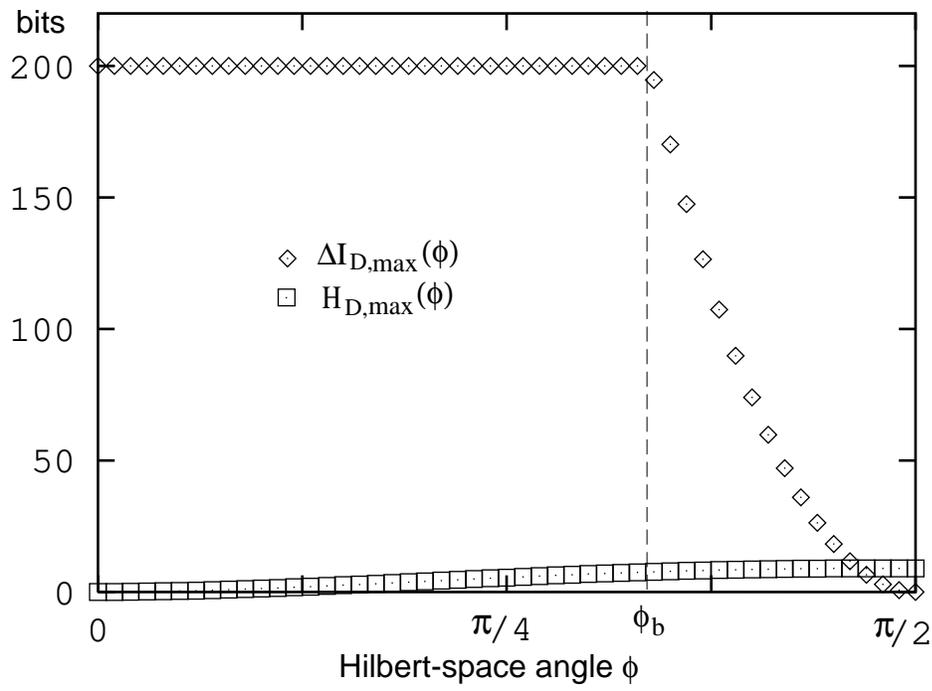
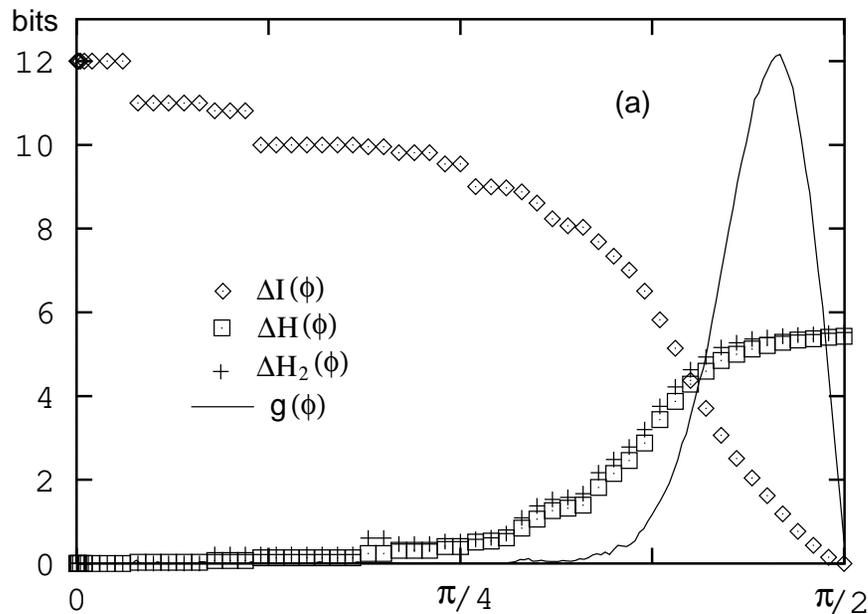
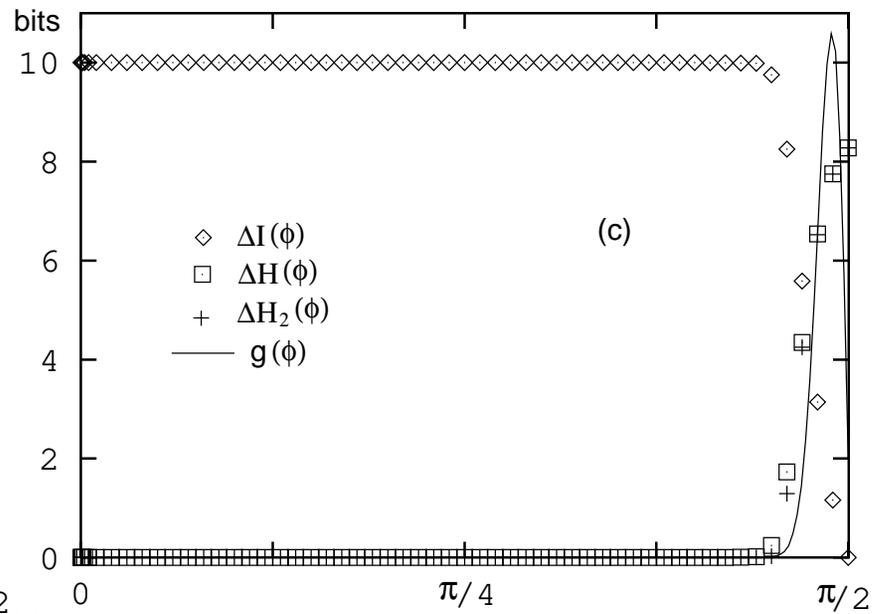
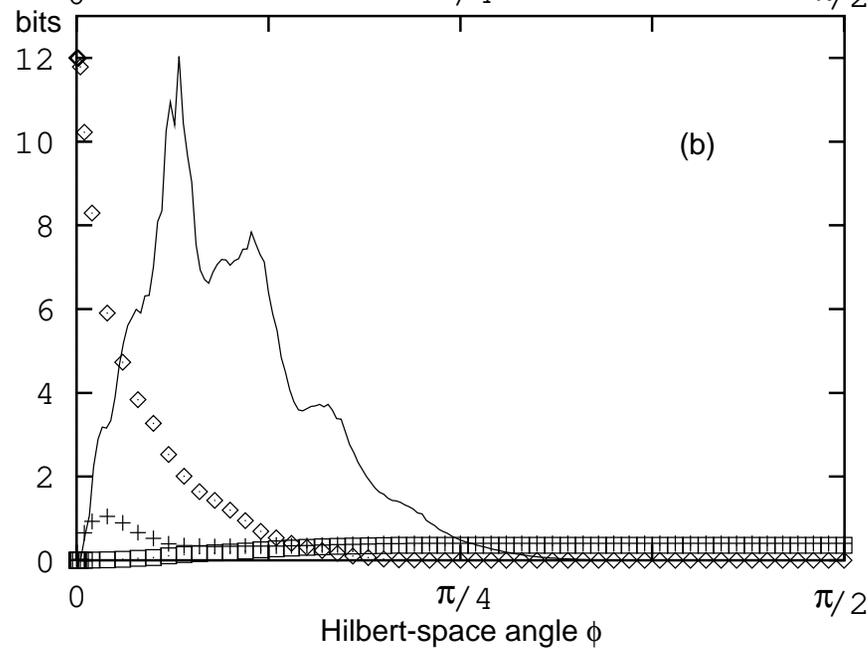
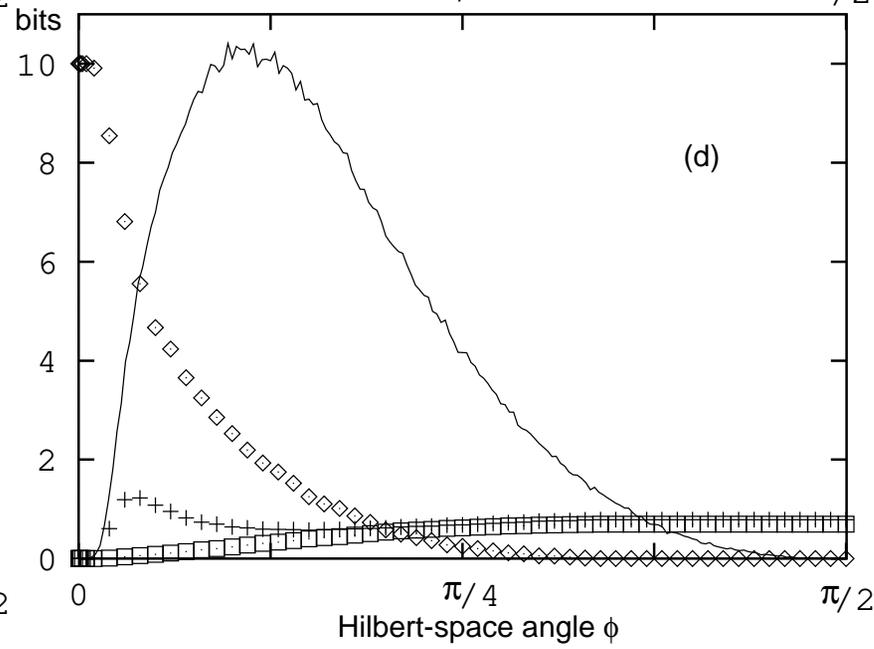
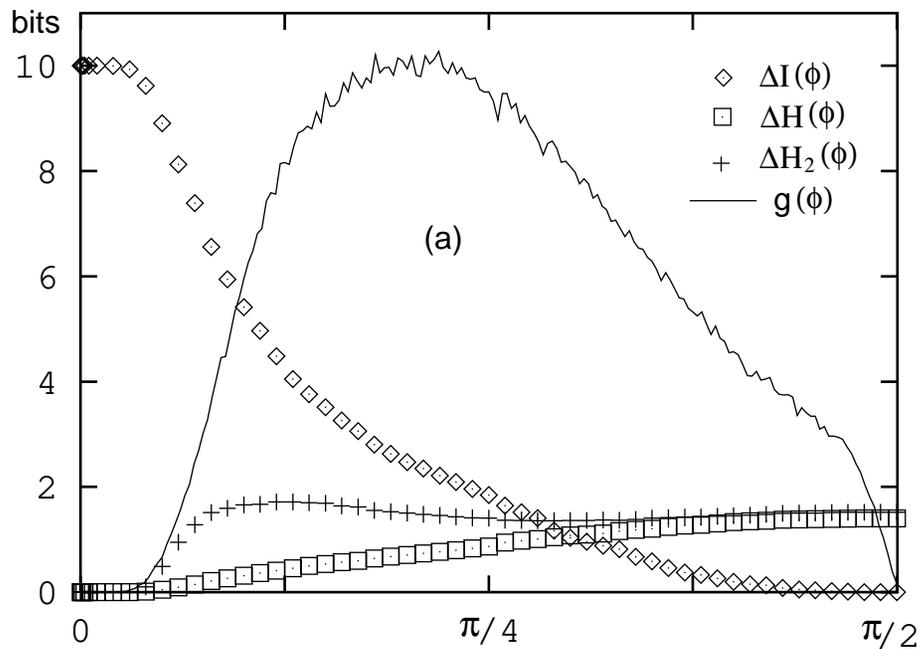
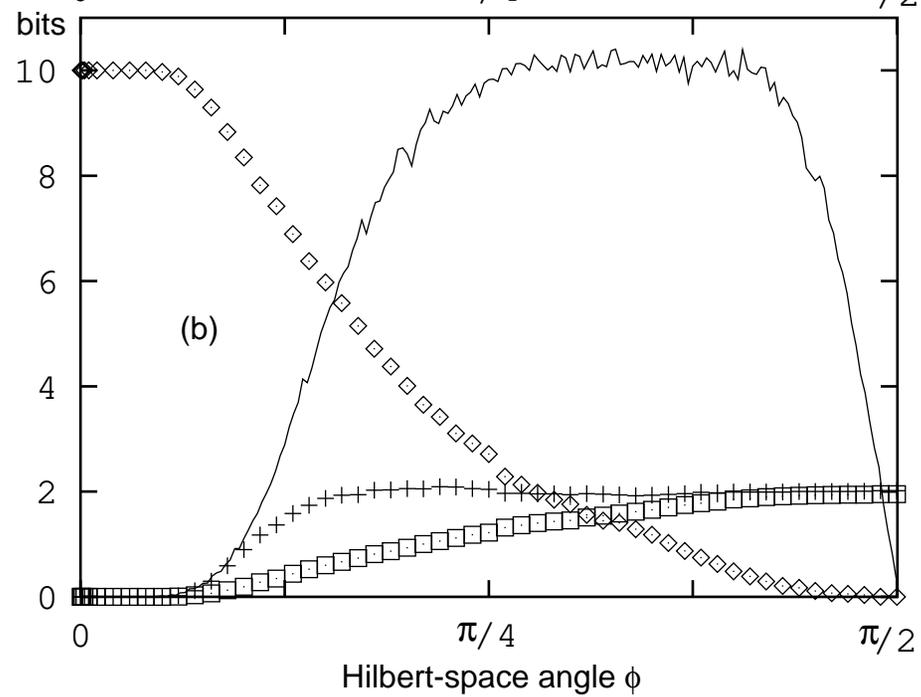
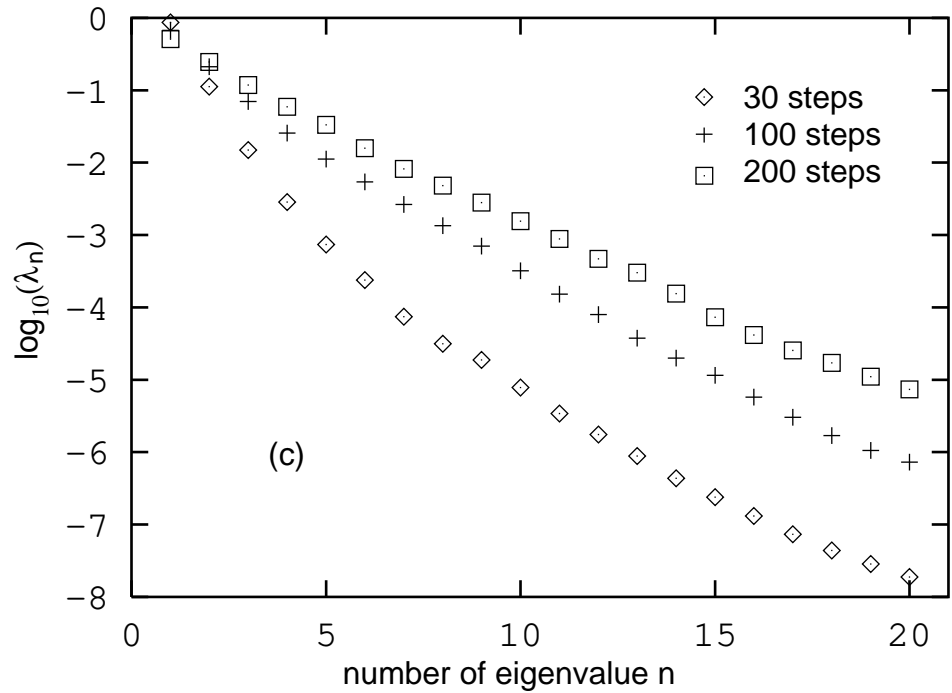
\end{document}